\documentclass[prc,aps,twocolumn,showpacs,preprintnumbers,nofootinbib,amsmath,amssymb]{revtex4}
\usepackage{graphicx} % Include figure files
\usepackage{psfig} % Include figure files
\usepackage{dcolumn}% Align table columns on decimal point
\begin{document}
\title{Strangeness dynamics and transverse pressure
in relativistic nucleus-nucleus collisions}

\author{E.~L.~Bratkovskaya}
 \affiliation{Institut f\"{u}r Theoretische Physik,
   J.W. Goethe Universit\"{a}t, 60054 Frankfurt, Germany}
\author{M. Bleicher}
 \affiliation{Institut f\"{u}r Theoretische Physik,
   J.W. Goethe Universit\"{a}t, 60054 Frankfurt, Germany}
\author{M. Reiter}
 \affiliation{Institut f\"{u}r Theoretische Physik,
   J.W. Goethe Universit\"{a}t, 60054 Frankfurt, Germany}
\author{S. Soff}
 \affiliation{Institut f\"{u}r Theoretische Physik,
   J.W. Goethe Universit\"{a}t, 60054 Frankfurt, Germany}
\author{H.~St\"ocker}
 \affiliation{Institut f\"{u}r Theoretische Physik,
   J.W. Goethe Universit\"{a}t, 60054 Frankfurt, Germany}
\author{M. van Leeuwen}
 \affiliation{NIKHEF, Amsterdam, Netherlands}
 \affiliation{Lawrence Berkeley National Laboratory, Berkeley, CA 94720, USA}
\author{S. A. Bass}
\affiliation{Department of Physics, Duke University, Durham, NC 27708, USA}
\affiliation{RIKEN BNL Research Center, Brookhaven National Laboratory,
        Upton, NY 11973, USA}
\author{W. Cassing}
 \affiliation{Institut f\"{u}r Theoretische Physik,
   Universit\"{a}t Giessen, 35392 Giessen, Germany}
%\date{ }

\begin{abstract}
We investigate hadron production as well as transverse hadron spectra
from proton-proton, proton-nucleus and nucleus-nucleus collisions from
2 $A\cdot$GeV to 21.3 $A\cdot$TeV within two independent transport
approaches (HSD and UrQMD) that are based on quark, diquark, string and
hadronic degrees of freedom.  The comparison to experimental data on
transverse mass spectra from $pp$, $pA$ and C+C (or Si+Si) reactions
shows the reliability of the transport models for light systems.  For
central Au+Au (Pb+Pb) collisions at bombarding energies above $\sim$ 5
A$\cdot$GeV, furthermore, the measured $K^{\pm}$ transverse mass
spectra have a larger inverse slope parameter than expected from the
default calculations.  We investigate various scenarios to explore
their potential effects on the $K^\pm$  spectra. In particular the
initial state Cronin effect is found to play a substantial role at top
SPS and RHIC energies.  However, the maximum in the $K^+/\pi^+$ ratio
at 20 to 30 A$\cdot$GeV  is missed by ~40\%  and the approximately
constant slope of the $K^\pm$ spectra at SPS energies is not reproduced
either.  Our systematic analysis suggests that the additional pressure
- as expected from lattice QCD calculations at finite quark chemical
potential $\mu_q$ and temperature $T$- should be generated by strong
interactions in the early pre-hadronic/partonic phase of central Au+Au
(Pb+Pb) collisions.
\end{abstract}


\maketitle

%**********************************************************************
\section{Introduction}

The phase transition from partonic degrees of freedom (quarks and
gluons) to interacting hadrons is a central topic of modern high-energy
physics \cite{QM02}. In order to understand the dynamics and relevant
scales of this transition laboratory experiments under controlled
conditions are presently performed with ultra-relativistic
nucleus-nucleus collisions.  Hadronic spectra and relative hadron
abundances from these experiments reflect  important aspects of the
dynamics in the hot and dense zone formed in the early phase of the
reaction. Here different quark flavors ($u,d,s,c,b$) - as constituents
of the final hadrons - provide valuable information on the early high
density phase due to the substantial different mass scales for the
light ($u,d$), strange ($s$) and charm ($c$) or bottom ($b$) quarks.
Furthermore, as has been proposed early by Rafelski and M\"uller
\cite{Rafelski} in particular the strangeness degree of freedom might
play an important role in distinguishing hadronic and partonic
dynamics.

Lattice QCD calculations at vanishing quark chemical potential and
finite temperature indicate critical energy densities for the formation
of a quark-gluon plasma (QGP) of $\sim$ 0.7-1 GeV/fm$^3$ \cite{Karsch},
which might already be achieved at Alternating Gradient Synchrotron
(AGS) energies of $\sim$ 10 $A\cdot$GeV for central Au+Au collisions
\cite{HORST,exita,Weber98}. Moreover, the recent lattice QCD
calculations at finite  quark chemical potential $\mu_q$ \cite{Fodor}
show a rapid increase of the thermodynamic pressure $P$ with
temperature above the critical temperature $T_c$ for a phase transition
(or cross-over) to the QGP. The problem is how to relate these numbers
to observables from nucleus-nucleus collisions.  Estimates based on the
Bjorken formula \cite{bjorken} for the energy density achieved in
central Au+Au collisions suggest that the critical energy density for
the formation of a QGP of $\sim$ 0.7-1 GeV/fm$^3$ is by far exceeded
during a few fm/c in the initial phase of Au+Au collisions at
Relativistic Heavy Ion Collider (RHIC) energies \cite{QM02}.
Accordingly, a phase transition (or cross-over) might already happen at
much lower bombarding energies.

Experimentally, relativistic nucleus-nucleus collisions have been
studied at beam energies from 0.1 to 2 $A\cdot$GeV at the
SchwerIonen-Synchrotron (SIS), from  2 to 11.6 $A\cdot$GeV at the  AGS
and  from 20 to 160 $A\cdot$GeV at the Super Proton Synchrotron (SPS)
\cite{NA49_new,NA49_T}. This program has been extended \cite{QM02} at
RHIC with Au+Au collisions at invariant energies $\sqrt{s}$ from $\sim
20$ to 200 GeV (equivalent energies in a fixed target experiment: 0.2
to 21.3 $A\cdot$TeV).

Whereas many previous experimental studies have focused on
longitudinal rapidity distributions for hadrons, the transverse mass
(or momentum) spectra of hadrons are presently in the center of interest.
On the one hand a significant suppression of high transverse momentum
hadrons in Au+Au collisions compared to $pp$ is observed at RHIC
energies of $\sqrt{s} =$ 200 GeV
\cite{survey,PHENIX_dAu,STAR_dAu,BRAHMS_dAu}.  This suppression is
often attributed to the energy loss of highly energetic particles in a
hot colored medium (QGP) \cite{Wang,Baier}.  In fact, the recent
observation by the PHENIX \cite{PHENIX_dAu}, STAR \cite{STAR_dAu} and
BRAHMS \cite{BRAHMS_dAu} collaborations, that a similar suppression is
not observed in d+Au interactions at midrapidities at the same energy,
supports this idea.  On the other hand, the measured transverse mass
($m_T=(p_T^2+m^2)^{1/2}$)
spectra of hadrons (heavier than pions)
\begin{eqnarray}
\label{slope}
\frac{1}{m_T} \frac{dN}{dm_T} \sim \exp\left(-\frac{m_T}{T}\right)
\end{eqnarray}
at  AGS, SPS and RHIC energies  show  a substantial {\it hardening}  in
central Au+Au collisions relative to $pp$ interactions (cf.
\cite{NA49_T,Goren}).  This hardening of the spectra (or increase of
the inverse slope parameter $T$ in (\ref{slope})) for low transverse mass
or momentum is commonly
attributed to strong collective flow, which is absent in the respective
$pp$ or $pA$ collisions.

The authors of \cite{SMES} have proposed to interpret the
approximately constant $K^\pm$ slopes above $\sim 30$ A$\cdot$GeV --
the 'step' -- as an indication for a phase transition.  This
interpretation is also based on a rather sharp maximum in the
$K^+/\pi^+$ ratio at $\sim$ 20 to 30 A$\cdot$GeV in central of
Pb+Pb collisions (the 'horn' \cite{SMES}).  However, the present
body of experimental data and our theoretical understanding does
not allow for a stringent conclusion based only on the 'step' and 'horn'
phenomena. In this paper we will
address some further theoretical issues, but would like to point
out that more experimental data, for example on collisions in the
energy range of $20 \leq \sqrt{s} \leq 200$~GeV and on weakly
interacting hadrons, will be needed to
clarify these issues from the experimental side.

We will demonstrate in this work, that neither the pressure needed to
generate a large collective flow -- to explain the hard slopes of the
$K^\pm$ spectra with a 'plateau' at SPS energies -- nor the
experimental sharp maximum in the $K^+/\pi^+$ ratio at $\sim$ 20 to 30
A$\cdot$GeV is produced in the present transport models by the
interactions of hadrons in the expansion phase of the hadronic
fireball. Some aspects of this work have been published in Refs.
\cite{Weber02,MT-prl} before; however, we here extend the previous
studies to RHIC energies and consider a wide variety of dynamical
scenarios.  In particular we will explore two distinct effects: i)
initial state Cronin enhancement and ii) heavy resonance formation
on the transverse mass spectra of pions and kaons at all bombarding
energies from lower AGS to RHIC energies.

In our studies we use two independent relativistic transport models
that employ hadronic and string degrees of freedom, i.e. UrQMD
\cite{UrQMD1,UrQMD2} and HSD \cite{Geiss,Cass99}. They take into
account the formation and multiple rescattering of hadrons and thus
dynamically describe the generation of pressure in the hadronic
expansion phase. This involves also interactions of leading pre-hadrons
that contain a valence quark (antiquark) from a primary 'hard'
collision (cf. Refs.  \cite{Weber02,Cass_Cron03}). In the parton
language, the two transport models employed do not include gluon
degrees of freedom explicitly nor gluon-gluon and gluon-quark
(antiquark) interactions.  We will use these models to test, whether
the description of nucleus-nucleus reactions in terms of pre-hadronic,
hadronic and string degrees of freedom and their interactions is
sufficient to explain the experimental data.

Our work is organized as follows: In Section II we  briefly recall the
ingredients of the HSD and UrQMD transport approaches and point out the
extensions incorporated in UrQMD (versions 2.0 and 2.1).  In Section III the
calculated excitation functions for pions, $K^\pm$ mesons and hyperons
for central Au+Au (or Pb+Pb) collisions from 2 to 21300 A$\cdot$GeV are
shown (in comparison to the experimental data) as well as selected
particle ratios. Section IV is devoted to transverse mass spectra of
pions and $K^\pm$ mesons from $pp$ to $pA$ and central $AA$ reactions
in the same energy range. In Section V we will discuss various
alternative scenarios within  the transport calculations to explore
their sensitivity to the hadron transverse mass spectra.  Section VI is
devoted to a study of the 'Cronin effect' in central  Au+Au collisions,
which we attribute to a 'pre-hadronic' initial state scattering effect
that also shows up in $pA$ reactions. Section VII includes a comparison
of the thermodynamic Lagrange parameters $T$ and $\mu_B$ extracted from
the UrQMD transport model in the central overlap regime of Au+Au
collisions  with the experimental systematics on chemical freeze-out
configurations in the $T,\mu_B$ plane as well as recent lattice QCD
results.  Section VIII closes this study with a summary and discussion
of open problems.

%**********************************************************************

\section{Transport models -- HSD and UrQMD}

We use two independent relativistic transport approaches for our study
that employ hadronic and string degrees of freedom, i.e. UrQMD
(Ultra-relativistic Quantum Molecular Dynamics) \cite{UrQMD1,UrQMD2}
and HSD (Hadron-String Dynamics) \cite{Geiss,Cass99,HSDcode} in order
to obtain an estimate of systematic uncertainties within the transport
models.  Both approaches take into account the formation and decay of
strings as well as the multiple rescattering of hadrons in the
expansion phase.  These models can be considered as appropriate
descriptions at least for the final stage of an ultrarelativistic
nucleus-nucleus collision, where the strongly interacting hadrons
dominate. The open question is, however, if these transport models also
incorporate the proper degrees of freedom in the very early phase of
heavy-ion collisions.  We stress that transport calculations allow for
a systematic study of the change in the dynamics from elementary
baryon-baryon or meson-baryon collisions to proton-nucleus reactions or
from peripheral to central nucleus-nucleus collisions in a unique way
without change in parameters.  This is of central importance since the
assumptions of thermal and chemical equilibrium -- as used in
hydrodynamical or statistical models \cite{BMJS} -- do not necessarily
hold in all these reactions and the transport studies allow to explore
the amount of (thermal or chemical) equilibrium reached
\cite{Bravina,Brat00,Ko_AMPT}.

The default UrQMD 1.3 transport approach  \cite{UrQMD1,UrQMD2} includes
all baryonic resonances up to  masses of 2 GeV as well as mesonic
resonances up to 1.9 GeV as tabulated by the Particle Data Group
\cite{PDG}. For hadronic continuum excitations a string model is used
with meson formation times of the order of 1-2~fm/c depending on the
momentum and energy of the created hadron. For the present study -- and
the particular aim to address also jet production and fragmentation at
RHIC energies -- we additionally have incorporated PYTHIA (v 6.1)
\cite{PYTHIA} in UrQMD 1.3 similar to HSD. This novel and extended
version we denote as UrQMD 2.0.

Moreover, to explore the physical consequences of higher mass states as
implemented in the Relativistic-Quantum-Molecular-Dynamics model (RQMD)
\cite{Sorge}, we have introduced in UrQMD additional high mass resonances
that are explicitly produced and propagated in {\em s-channel} processes
with invariant masses up to $\sqrt s<3$~GeV. We do not assume these
states to be excited in baryon-baryon collisions. In practical terms:
If the mass of the produced mesonic or baryonic resonance is above
$\sim 2$~GeV\footnote{The explicit value depends on the type of
resonance.}, i.e in the continuum, the formed resonance state is not
fragmented as a string object like in UrQMD 1.3 or UrQMD 2.0. We recall
that a string of invariant mass 2 $\leq M \leq$ 3 GeV essentially
decays to a few pions, a vector meson + pions etc.  Instead, this high
mass state is treated as a quasi-particle that decays according to the
branching ratios of the resonances with a mass of 1.9 to 2.0 GeV (as
implemented in UrQMD). We recall that baryon resonances $\sim$ 1.9 GeV
dominantly decay to the $\pi+N$ channel and with some lower probability
to the $\rho (\omega)+N$ final state. In this way light meson emission
is suppressed by about 25\% compared to a string of the same invariant
mass. We note, that this recipe is still in line with the available
$\pi p\to \pi X$ data on pion multiplicities. Furthermore, since the
high mass resonances decay isotropically in their rest frame, this
leads to an enhancement of transverse flow (cf. Section V.E).  
Furthermore, we use isotropic angular distributions for elastic 
meson-baryon interactions below $\sqrt{s}_{mB}=6$ GeV.
In this work we will denote this scenario as UrQMD 2.1.

In the HSD approach nucleons, $\Delta$'s, N$^*$(1440), N$^*$(1535),
$\Lambda$, $\Sigma$ and $\Sigma^*$ hyperons, $\Xi$'s, $\Xi^*$'s and
$\Omega$'s  as well as their antiparticles are included on the baryonic
side whereas the $0^-$ and $1^-$ octet states are included in the
mesonic sector. Inelastic hadron--hadron collisions with energies above
$\sqrt s\simeq$ 2.6~GeV are described by the FRITIOF model \cite{LUND}
(incorporating PYTHIA v 5.5 with JETSET v 7.3 for the production and
fragmentation of jets \cite{PYTHIA}) whereas low energy hadron--hadron
collisions are modeled in line with experimental cross sections
\cite{Ehehalt,Geiss,Cass99}.

Both transport approaches reproduce the nucleon-nucleon, meson-nucleon
and meson-meson cross section data in a wide kinematic range
\cite{UrQMD1,UrQMD2,Weber02}. Therefore, we expect that the 'hadronic'
scattering mechanisms are sufficiently under control. However, in
nucleus-nucleus reactions further elastic and inelastic processes on
the parton level might occur that -- beyond the interactions of
'leading' quarks/diquarks -- are not included in the studies below. We
mention that (pQCD-like) partonic scattering is incorporated e.g. in the
multi-phase transport model (AMPT) \cite{Ko_AMPT}, which is currently
employed from upper SPS to RHIC energies, but is not suitable for lower
energies.

%**********************************************************************

\section{Excitation functions}

Whereas the underlying concepts of the transport theoretical models are
very similar, the actual realizations differ considerably.
A systematic analysis of results from both models and
experimental data for central nucleus-nucleus collisions from 2 to 160
$A\cdot$GeV has shown that the 'longitudinal' rapidity distributions of
protons, pions, kaons, antikaons and hyperons are quite similar in both
models and in  reasonable agreement with available data \cite{Weber02}.
We here extend the latter studies to RHIC energies, but focus on
total and midrapidity hadron yields as well as particle ratios.
For completeness we include the results from Ref. \cite{Weber02}
in Figs. \ref{Fig_yield}-\ref{Fig_rat}.

\subsection{Pion and strange particle yields}

\begin{figure*}[!]
%\phantom{a}\vspace*{-1.5cm}
\centerline{\psfig{figure=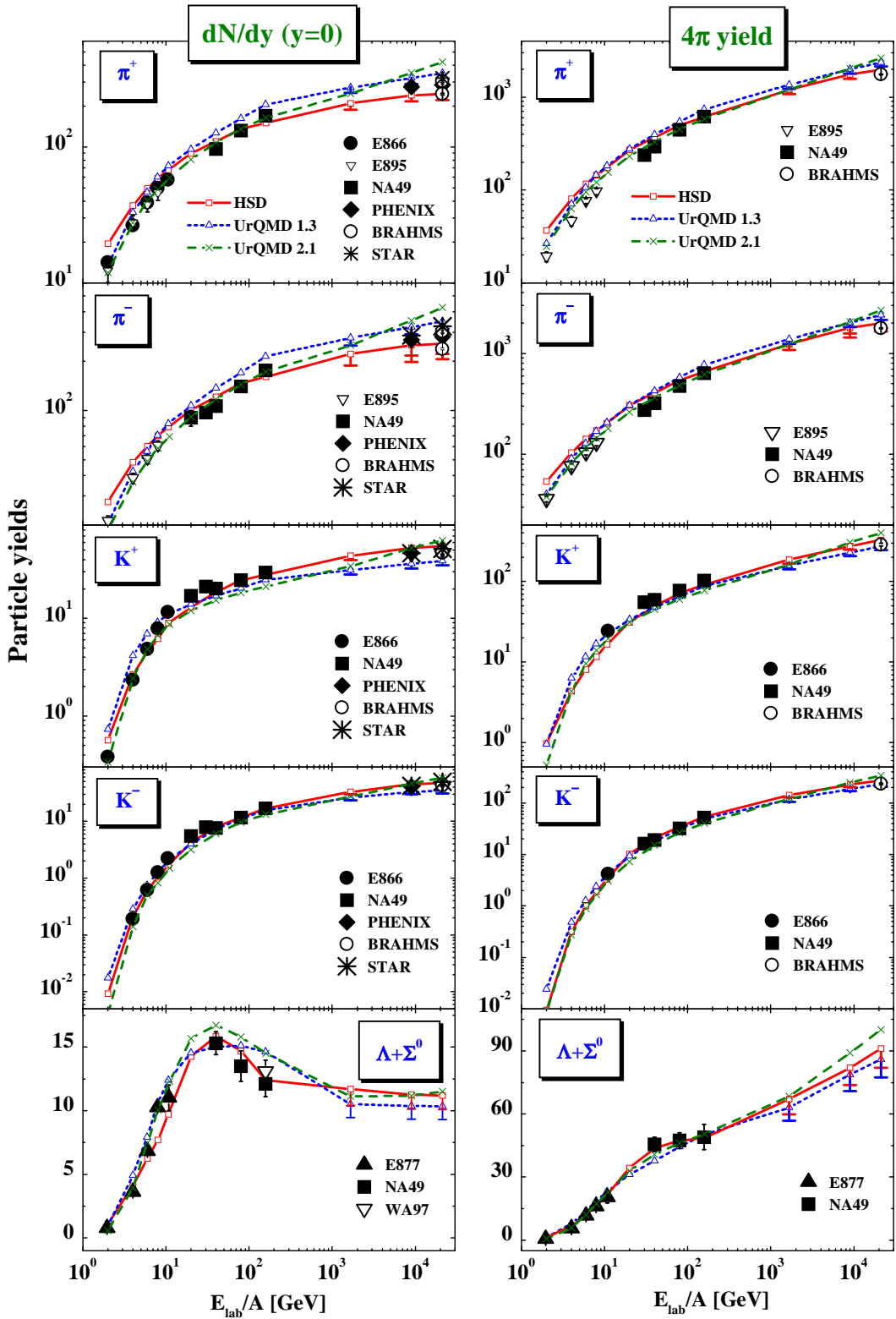,width=14cm}}
\caption{(Color online) The excitation function of $\pi^+, \pi^-, K^+,
K^-$ and $\Lambda+\Sigma^0$ yields from 5\% central (AGS energies, SPS
at 160 A$\cdot$GeV and at RHIC energies), 7\% central (20, 30, 40 and
80 A$\cdot$GeV), 10\% central for $\Lambda+\Sigma^0$ at 160 A$\cdot$GeV
Au+Au (AGS and RHIC) or Pb+Pb (SPS) collisions in comparison to the
experimental data from Refs.  \protect\cite{E866E917,E895,E891Lam}
(AGS), \protect\cite{NA49_new,NA49_Lam,Antiori,NA49_QM04} (SPS) and
\protect\cite{BRAHMS,PHENIX,STAR} (RHIC) for midrapidity (left column)
and rapidity integrated yields (right column).  The solid lines with
open squares show the results from HSD whereas the dashed lines with
open triangles and the long dashed lines with crosses indicate the
results from UrQMD 1.3 and UrQMD 2.1, respectively.  The lower
theoretical errorbars at RHIC energies correspond to the yields for
10\% central events.}
\label{Fig_yield}
\end{figure*}

\begin{figure*}[!]
\centerline{\psfig{figure=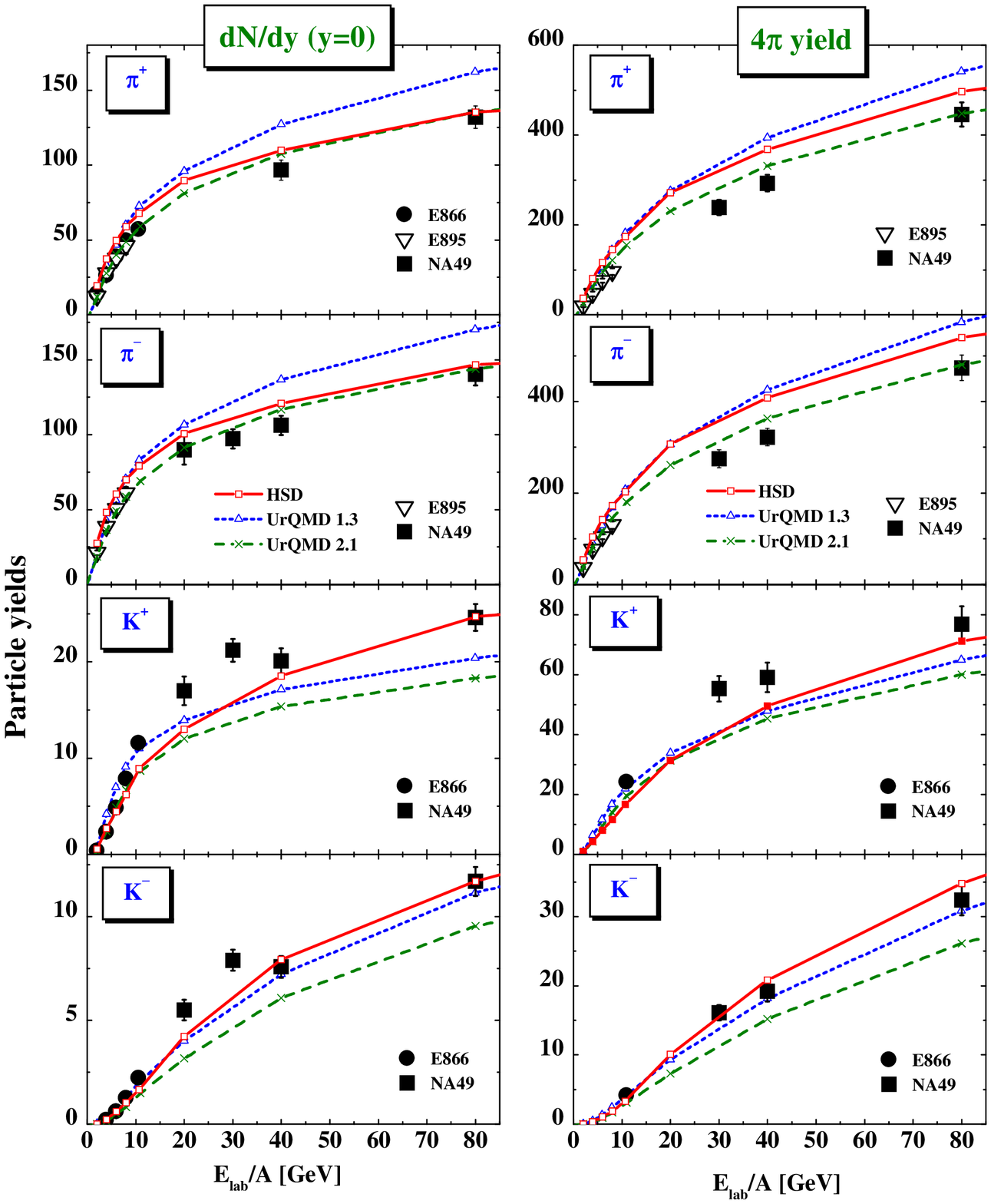,width=14cm}}
\caption{(Color online) The excitation function of $\pi^+, \pi^-, K^+$
and $K^-$ mesons (as in Fig.\protect\ref{Fig_yield}), but
up to 80 A GeV  on a linear scale.}
\label{Fig_yieldb}
\end{figure*}

Figure \ref{Fig_yield} shows the excitation function of $\pi^+, \pi^-,
K^+, K^-$ and $\Lambda+\Sigma^0$ yields from 5\% central (AGS energies,
SPS at 160 A$\cdot$GeV and at RHIC energies), 7\% central (20, 30, 40
and 80 A$\cdot$GeV), 10\% central for $\Lambda+\Sigma^0$ at 160
A$\cdot$GeV Au+Au (AGS and RHIC) or Pb+Pb (SPS) collisions in
comparison to the experimental data from Refs.
\cite{E866E917,E895,E891Lam} (AGS),
\cite{NA49_new,NA49_Lam,Antiori,NA49_QM04} (SPS) and
\cite{BRAHMS,PHENIX,STAR} (RHIC) for midrapidity (left column) and
rapidity integrated yields (right column).  The solid lines with open
squares show the results from HSD, the dashed lines with open triangles
indicate the calculations with UrQMD 1.3 and the long dashed lines with
crosses depict the UrQMD 2.1 results.

We note in passing, that the corrections to the pion and kaon yields
due to the weak decay of resonances have not been included in our
calculations in line with  the experimental analysis of the NA49
Collaboration (cf. Refs.  \cite{NA49_new,NA49_T}). However, the
experimental subtraction of week decays is problematic and might have
an impact on the  measured kaon to pion ratio as pointed out in Ref.
\cite{Andronic}. This uncertainty has to be kept in mind throughout the
following analysis.

As already demonstrated in Ref. \cite{Weber02} the agreement between
the transport models HSD (solid lines) and UrQMD 1.3 (dashed lines) is
rather good as well as with the data available.  The exception are the
pion yields at the highest AGS energy and lower SPS energies, which are
overestimated by both approaches (cf.  \cite{Weber02}). This is
demonstrated in more detail in Fig.  \ref{Fig_yieldb} where the
$\pi^\pm$ and $K^\pm$ yields (as in Fig. \ref{Fig_yield}) are displayed
in the energy range up to 80 A GeV  on a linear scale. The
overestimation of the pion yields (in 4$\pi$) in this energy range is
slightly more pronounced for UrQMD 1.3 than for HSD especially at the
higher SPS energy. On the other hand, both transport models fall
slightly low in the $K^+$ yields at 30 and 40 A GeV.  The results from
the novel version UrQMD 2.1 (long dashed lines) differ from UrQMD 1.3
essentially in the following aspect:  The pion multiplicity at AGS and
SPS energies is reduced by about 25\% (cf. Section II) due to the
specific decay scheme of higher mass resonances (and is now in better
agreement with the data). On the other hand, UrQMD 2.1 gives a lower
amount of kaons and especially antikaons (in $4\pi$) than the 'default'
HSD and UrQMD 1.3 approaches.

At RHIC energies the UrQMD 1.3 calculations show a $\sim$ 30\% higher
pion yield at midrapidity than HSD while the differences between the
4$\pi$ integrated yields are of the order of 10\%, both between the
models and compared to the data.  The implementation of  higher mass
states in UrQMD 2.1 (cf. Section II) -- together with the jet
production described by PYTHIA -- leads to an overestimation of pions,
kaons and hyperons at RHIC energies.

Note, that the maximum in the $\Lambda + \Sigma^0$ yield at midrapidity
(lower left part of Fig. \ref{Fig_yield}) for $\sim$ 30 A$\cdot$GeV is
reproduced by both transport approaches.  However, this maximum
disappears in the 4$\pi$ integrated yields in line with the data (cf.
lower right part of Fig. \ref{Fig_yield}).  Within the transport models
the  maximum in the $\Lambda+\Sigma^0$ yield at midrapidity can be
attributed to a large extent to a change from baryon dominated dynamics
at AGS energies to meson dominated dynamics at the higher SPS energies.

\begin{figure*}[!]
%\phantom{a}\vspace*{-1.5cm}
\centerline{\psfig{figure=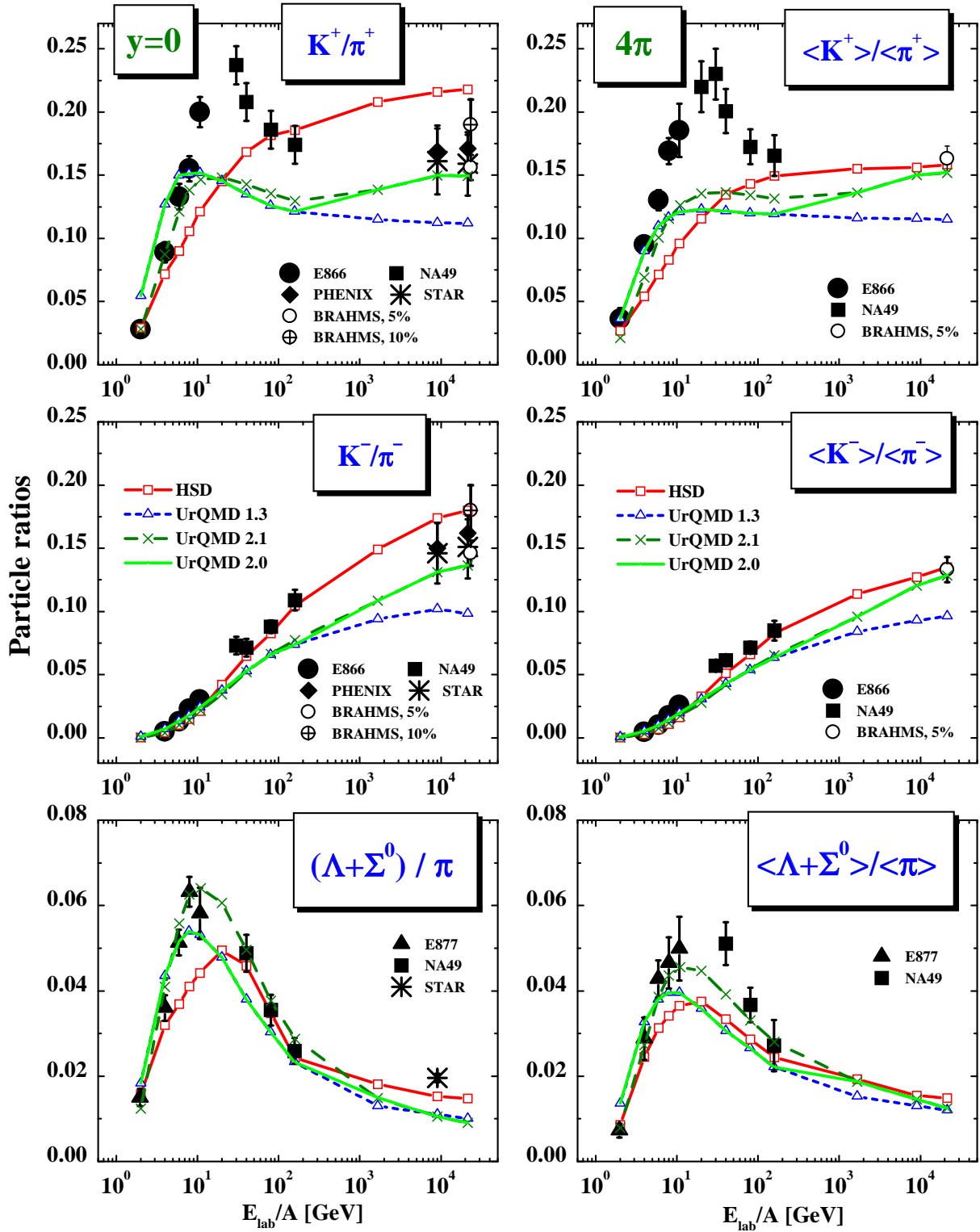,width=16cm}}
\caption{(Color online) The excitation function of $K^+/\pi^+, K^-/\pi^-$
and $(\Lambda+\Sigma^0)/\pi$ ratios from 5\% central (AGS energies,
SPS at 160 A$\cdot$GeV and at RHIC energies), 7\% central (20, 30, 40
and 80 A$\cdot$GeV), 10\% central for $\Lambda+\Sigma^0$ at 160 A$\cdot$GeV
Au+Au (AGS and RHIC) or Pb+Pb (SPS) collisions in comparison to the
experimental data from Refs. \protect\cite{E866E917,E891Lam}
(AGS), \protect\cite{NA49_new,NA49_Lam,Antiori,NA49_QM04} (SPS) and
\protect\cite{BRAHMS,PHENIX,STAR} (RHIC) for midrapidity (left column)
and rapidity integrated yields (right column).  The solid lines with
open squares show the results from HSD whereas the dashed lines with
open triangles and the long dashed lines with crosses
indicate the results from UrQMD 1.3 and UrQMD 2.1, respectively.}
\label{Fig_rat}
\end{figure*}

\begin{figure*}[!]
%\phantom{a}\vspace*{-1.5cm}
\centerline{\psfig{figure=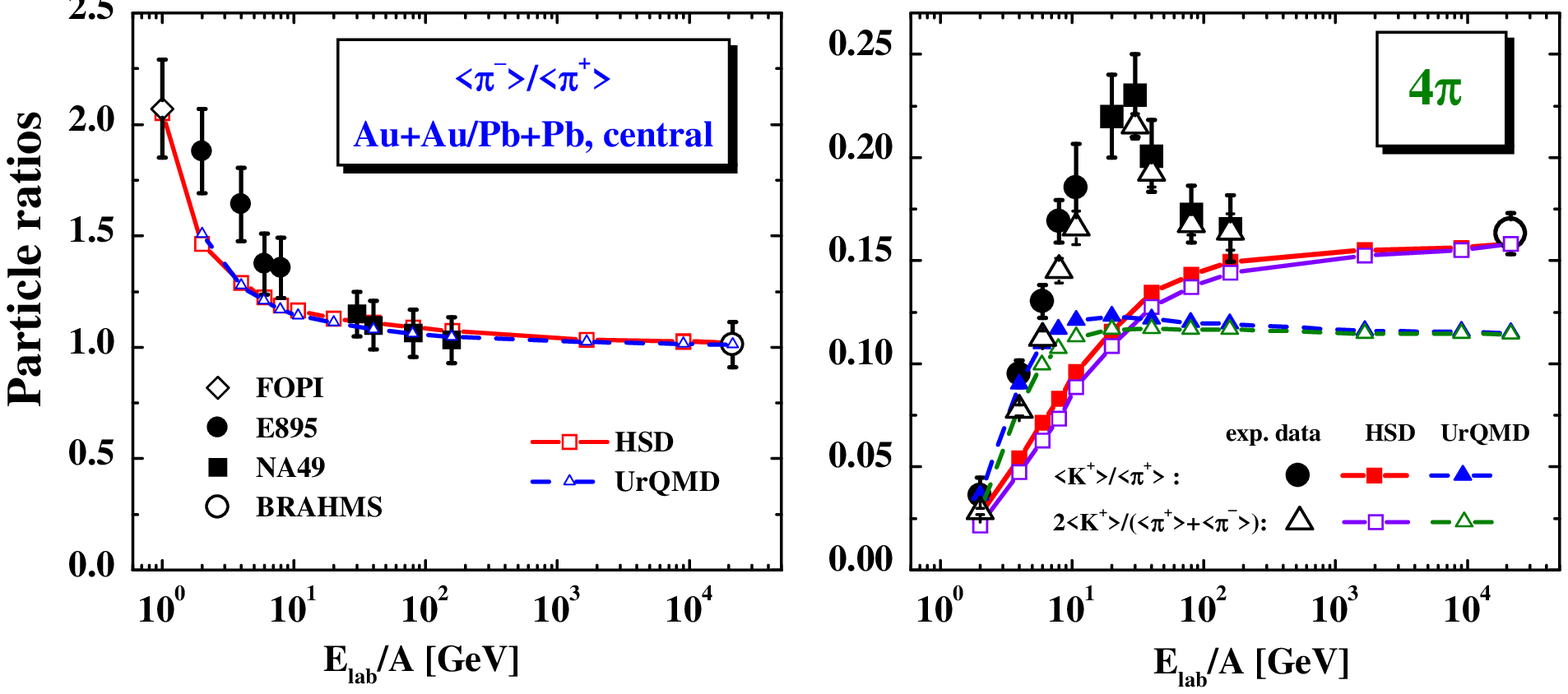,width=15cm}}
\caption{(Color online) The excitation function of the $\langle
\pi^-\rangle/\langle \pi^+\rangle$ (l.h.s.) and $\langle
K^+\rangle/\langle \pi\rangle$ ratios (r.h.s.) from central Au+Au (or
Pb+Pb) collisions.  Left part:  the HSD results are shown by the solid
line with open squares, whereas the UrQMD 1.3 calculations are
displayed in terms of the solid line with open triangles. The
experimental data have been taken from Refs.
\protect\cite{FOPI,E895,NA49_new,BRAHMS}.  Right part: the solid line
with solid squares (triangles) (cf. Fig.  \protect\ref{Fig_rat})
corresponds to the HSD (UrQMD 1.3) calculations for $\langle
K^+\rangle/\langle \pi^+\rangle$ (as well as the experimental data from
Refs.  \protect\cite{E866E917,NA49_new,NA49_QM04,BRAHMS}) while the solid
line with open squares (triangles) shows the calculated $2 \langle
K^+\rangle/(\langle \pi^+\rangle+\langle \pi^-\rangle)$ ratio in
comparison to the experimental data (open triangles) from Ref.
\protect\cite{MarcoSQM03}. }
\label{Fig_ratpiav}
\end{figure*}

\subsection{$K^+/\pi^+, K^-/\pi^-$ and $(\Lambda+\Sigma^0)/\pi$
ratios}

In Fig. \ref{Fig_rat} we present the excitation function of the
particle ratios $K^+/\pi^+, K^-/\pi^-$ and $(\Lambda+\Sigma^0)/\pi$
from 5\% central (AGS energies, SPS at 160 A$\cdot$GeV and at RHIC
energies), 7\% central (20, 30, 40 and 80 A$\cdot$GeV), 10\% central
for $\Lambda+\Sigma^0$ at 160 A$\cdot$GeV Au+Au (AGS and RHIC) or Pb+Pb
(SPS) collisions in comparison to the experimental data from Refs.
\cite{E866E917,E895,E891Lam} (AGS), \cite{NA49_new,NA49_Lam,Antior,NA49_QM04}
(SPS) and \cite{BRAHMS,PHENIX,STAR} (RHIC) for midrapidity yields (left
column) and rapidity integrated yields (right column).  The solid lines
with open squares show the results from HSD, the dashed lines with open
triangles indicate the UrQMD 1.3 calculations whereas the long dashed
lines with crosses correspond to the UrQMD 2.1 results.  The deviations
between the transport models and the data are most pronounced for the
midrapidity ratios (left column) since the ratios are very sensitive to
actual rapidity spectra. The $K^+/\pi^+$ ratio in UrQMD 1.3 shows a
maximum at $\sim$ 8 A$\cdot$GeV and then drops to a roughly constant
ratio of 0.12 at top SPS and RHIC energies. UrQMD 2.1 follows the
results from UrQMD 1.3 at AGS and SPS energies but then increases
slightly up to RHIC energies.  In case of HSD a continuously rising
ratio with bombarding energy is found for the midrapidity ratios. This
rise up to the full RHIC energy is essentially due to an
underestimation of the pion yield at midrapidity (cf. Fig. 1 in Ref.
\cite{Brat03}). The 4$\pi$ ratio in HSD, which is approximately
constant from top SPS to RHIC energies, is larger than the ratio from
UrQMD 1.3 due to the lower amount of pion production and a slightly
higher $K^+$ yield (cf. Fig.  \ref{Fig_yield}).  We note that the lower
amount of pions in HSD is essentially due to an energy-density cut (in
the local rest frame), which does not allow  hadron formation when is
energy density is above 1 GeV/fm$^3$.

Despite this seemingly good description of the separate particle
yields, the experimentally observed maximum in the $K^+/\pi^+$ ratio at
20-30 A$\cdot$GeV is not reproduced in the models.  Due to strangeness
conservation, the same arguments hold for the $(\Lambda +\Sigma^0)/\pi$
ratio, where the pronounced experimental maxima are underestimated due
to the excess of pions in the transport models at top AGS energies (for
HSD) and above $\sim$ 5 A$\cdot$GeV (for UrQMD 1.3). Here UrQMD 2.1
performs better in the ratios due to the lower amount of pions produced
by the high resonance decays.  Since the $K^-$ yields are well
described by all approaches (cf. Fig. \ref{Fig_yield}) the deviations
in the $K^-/\pi^-$ ratios at SPS and RHIC energies in UrQMD 1.3 indeed
can be traced back to the excess of pions. This suggests that the
entropy production, which is reflected in the pion to baryon ratio at
these energies, is overestimated in UrQMD 1.3 (and HSD) above $\sim$ 5
A$\cdot$GeV in central Au+Au collisions. Note, that in UrQMD 2.1 the
entropy production is reduced by the generation of high mass states in
$s$-channels, which essentially decay by emitting a lower amount of
pions.

It has been discussed in the community \cite{Fischer}, that the maximum
in the $K^+/\pi^+$ ratio might be due to isospin effects in Au+Au
(Pb+Pb) collisions arising from a different rapidity dependence of $u$
and $d$ quarks in $pp$, $pn$ and $nn$ collisions. We here address this
issue and show in Fig. \ref{Fig_ratpiav}  the influence of isospin
effects on the kaon to pion ratio. The left part of Fig.
\ref{Fig_ratpiav} displays the excitation function of the $\langle
\pi^-\rangle/\langle \pi^+\rangle$ ratio from central Au+Au (or Pb+Pb)
collisions calculated within HSD (solid line with open squares) and
UrQMD 1.3 (solid line with open triangles).  The right hand panel shows
both the isospin-sensitive $\langle K^+\rangle/\langle \pi^+\rangle$
ratio (open symbols) and the (approximately) isospin-insensitive $2
\langle K^+\rangle/(\langle \pi^+\rangle+\langle \pi^-\rangle)$ ratio
(solid symbols). The HSD results are shown as squares connected by
solid lines while the UrQMD 1.3 results are represented as  dashed
lines with triangles. The larger solid symbols indicate the
experimental results for $\langle K^+\rangle/\langle \pi^+\rangle$ from
Refs. \cite{E866E917,NA49_new,NA49_QM04,BRAHMS} and the large open triangles
correspond to the ratio $2 \langle K^+\rangle/(\langle
\pi^+\rangle+\langle \pi^-\rangle)$ from Ref.  \cite{MarcoSQM03}.
As seen from Fig. \ref{Fig_ratpiav}, the calculated (as well as
experimentally measured) isospin-insensitive ratio $2 \langle
K^+\rangle/(\langle \pi^+\rangle+\langle \pi^-\rangle)$ is only
slightly lower than the isospin-sensitive $\langle K^+\rangle/\langle
\pi^+\rangle$ ratio.  This result is due to the approximately constant
$\langle \pi^-\rangle/\langle \pi^+\rangle$ ratio above
$\sim 6$~A$\cdot$GeV (l.h.s. of Fig.  \ref{Fig_ratpiav}) in the
transport calculations as well as in the data.  We thus conclude that
isospin effects cannot be responsible for the maximum in the $\langle
K^+\rangle/\langle \pi\rangle^+$ ratio (r.h.s. of Fig.
\ref{Fig_ratpiav}).

%**********************************************************************

\section{Transverse mass spectra}

\begin{figure*}[!]
%\phantom{a}\vspace*{-1.5cm}
\centerline{\psfig{figure=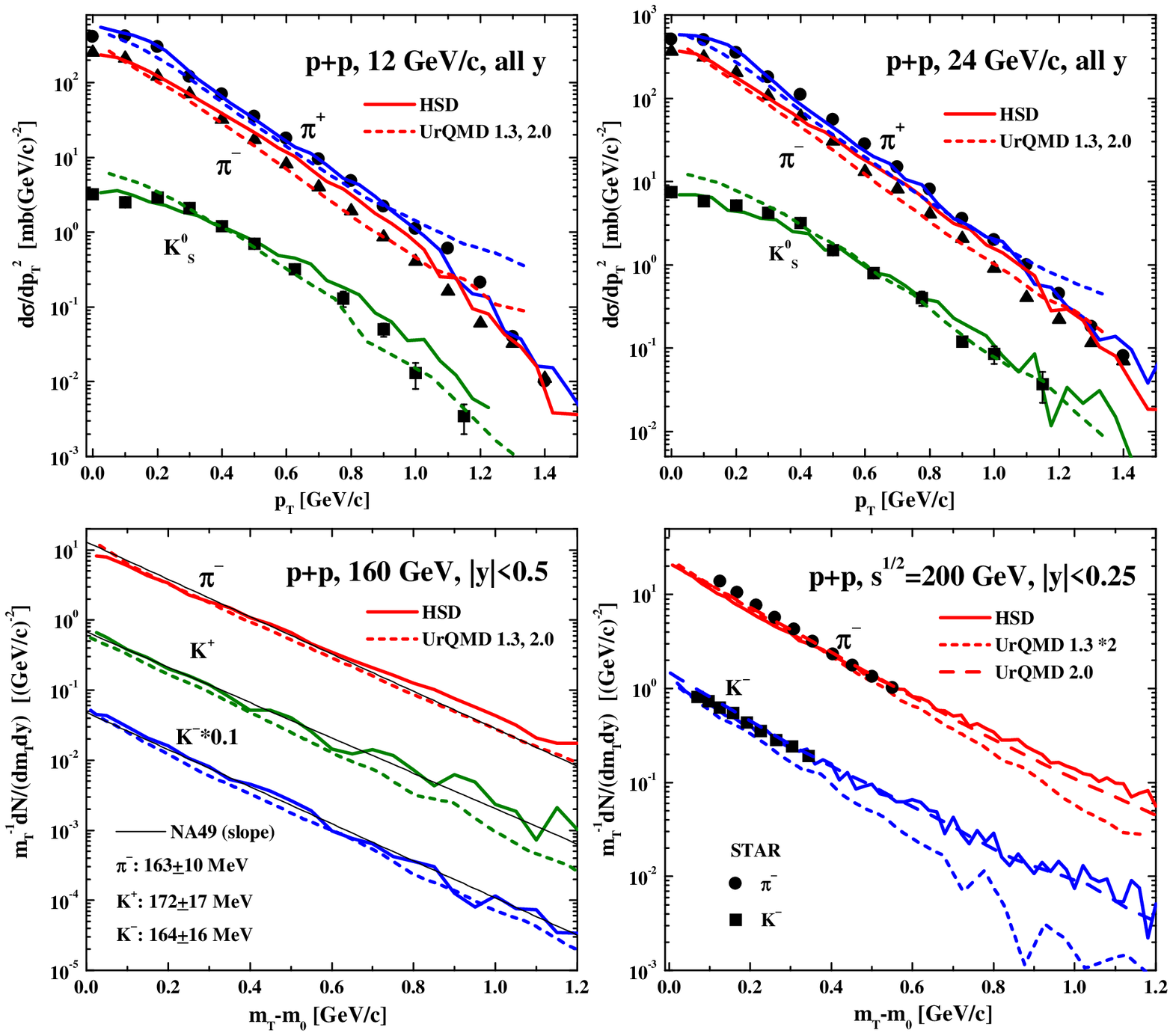,width=13.5cm}}
\caption{(Color online)
Upper part:
comparison of the differential cross sections for $\pi^+, \pi^-$ and
$K_S^0$ from $pp$ reactions at 12 and 24 GeV$/c$ from HSD (solid lines)
and UrQMD 1.3 (=UrQMD 2.0) (dashed lines) with the data from Ref.
\protect\cite{Blobel74}.
Lower part (left): the transverse mass spectra at midrapidity
for $\pi^-, K^+$ and $K^- (\times 0.1)$ from $pp$ reactions
at a bombarding energy 160 GeV from HSD (solid lines) and UrQMD
1.3 (dashed lines). The thin lines correspond to fits of the
experimental slope parameters $163\pm 10$ MeV for $\pi^-$, $172\pm 17$ MeV
for $K^+$ and $164\pm 16$ MeV for $K^-$ from Ref. \protect\cite{NA49_CCSi}.
Lower part (right):
the transverse mass spectra at midrapidity for $\pi^-$ and $K^-$ from
$pp$ reactions at $\sqrt{s}=200$ GeV from HSD (solid lines)  and
UrQMD 1.3 (dashed lines, multiplied by a factor 2) as well as UrQMD 2.0
(=UrQMD 2.1) (long dashed lines). The full symbols indicate the data
from the STAR Collaboration \protect\cite{STAR}.}
\label{Fig_mtpp}
\end{figure*}

Despite the relatively good agreement of data and transport
calculations in the longitudinal direction \cite{Weber02}, significant
deviations are found in the transverse direction  as pointed out in
Ref. \cite{MT-prl} before. This demonstrates that longitudinal rapidity
distributions are less sensitive to the hadronic interaction channels
employed in the transport models.  We here present a more complete and
systematic analysis up to the full RHIC energy of $\sqrt{s}$ = 200~GeV
and investigate alternative scenarios.  We start with benchmark tests
for the transport calculations for $pp$ and $pA$ reactions.

\subsection{$pp$ and $pA$ reactions}

\begin{figure*}[!]
%\phantom{a}\vspace*{-1.5cm}
\centerline{\psfig{figure=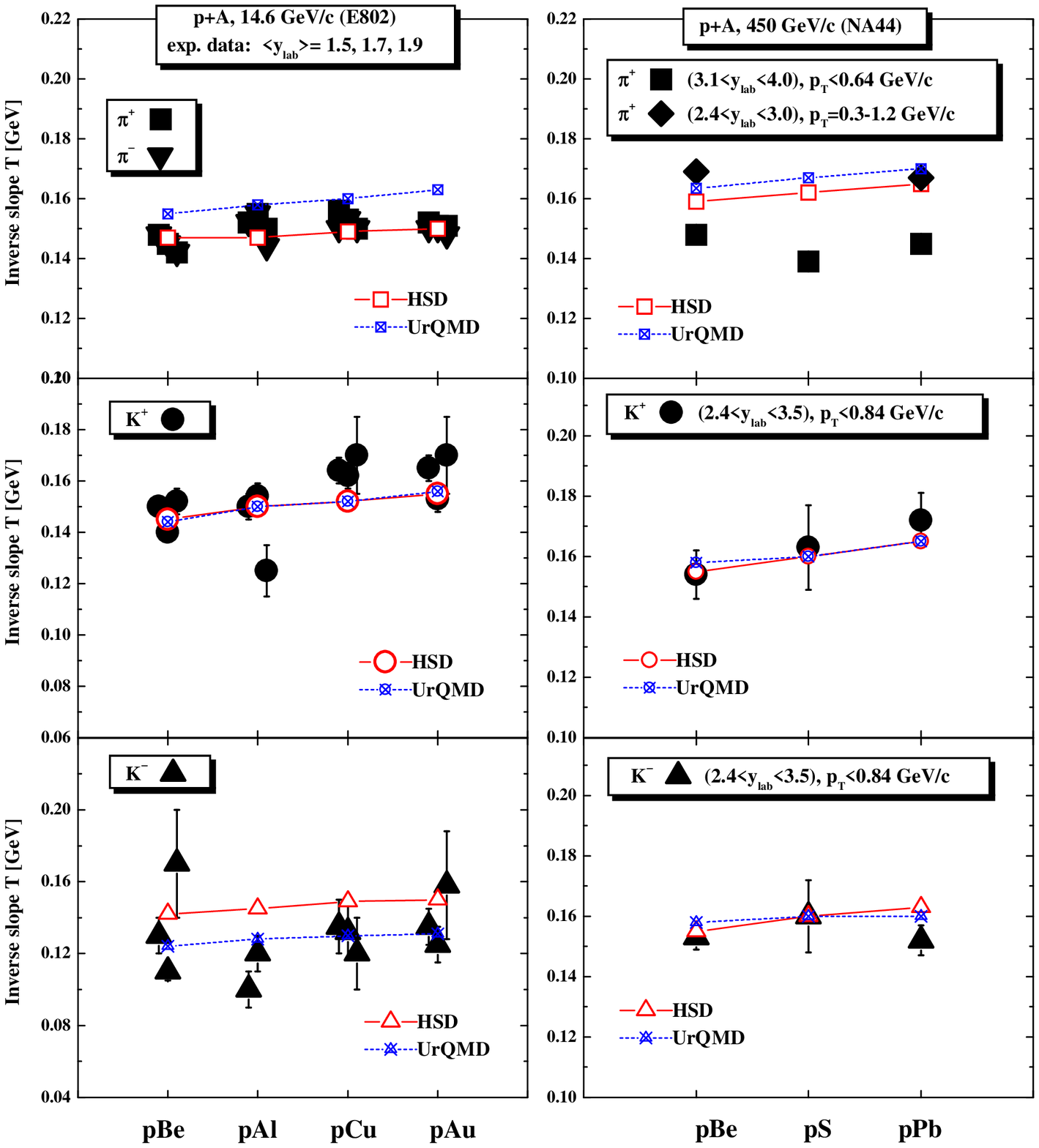,width=12.8cm}}
\caption{(Color online) Inverse slope parameters $T$ for $\pi^\pm, K^+$ and
$K^-$ at midrapidity from $pA$ reactions at 14.6 GeV$/c$ ($A=$Be, Al, Cu, Au)
 -- left part and at 450 GeV$/c$ ($A=$Be, S, Pb) -- right part, from
HSD (open symbols connected by solid lines) and UrQMD 1.3(2.0)
(open symbols with crosses connected by dashed lines).
The full symbols in the left part correspond to the midrapidity
data ($\langle y_{lab}\rangle = 1.5, 1.7, 1.9$) from the E802 Collaboration
\protect\cite{E802_pA},
in the right part to the NA44 data \protect\cite{NA44_pA}
at $2.4\le y_{lab}\le 3.5, \ p_T\le 0.84$ GeV$/c$ for $K^+, K^-$ and
at $2.4\le y_{lab}\le 3.0, \ p_T =0.3\div 1.2$ GeV$/c$ (full diamonds)
and $3.1\le y_{lab}\le 4.0, \ p_T\le 0.64$ GeV$/c$ (full squares)
for $\pi^+$. }
\label{Fig_tpa}
\end{figure*}

Our results for the transverse mass (or momentum) spectra of pions and
kaons from $pp$ reactions are presented in Fig. \ref{Fig_mtpp}. The
upper part includes the comparison of the differential cross sections
for $\pi^+, \pi^-$ and $K_S^0$ from $pp$ reactions at 12 and 24 GeV$/c$
from HSD (solid lines) and UrQMD 1.3 (dashed lines) with the data from Ref.
\cite{Blobel74}.  The lower left part of Fig. \ref{Fig_mtpp} shows the
transverse mass spectra at midrapidity for $\pi^-, K^+$ and $K^-
(\times 0.1)$ from $pp$ reactions at a bombarding energy of 160 GeV
from HSD (solid lines) and UrQMD 1.3 (dashed lines). The thin lines here
correspond to fits with the experimental slope parameters $163\pm 10$
MeV for $\pi^-$, $172\pm 17$ MeV for $K^+$ and $164\pm 16$ MeV for
$K^-$ from Ref. \cite{NA49_CCSi}.  The lower right part of Fig.
\ref{Fig_mtpp} displays the transverse mass spectra at midrapidity for
$\pi^-$ and $K^-$ from $pp$ reactions at $\sqrt{s}=200$ GeV from HSD
(solid lines), UrQMD 1.3 (dashed lines, multiplied by a  factor 2),
and UrQMD 2.0 (2.1) (long dashed lines).
The full symbols, furthermore, stand for
the data from the STAR Collaboration \cite{STAR}.
The description of the transverse mass spectra for pions and kaons
is reasonably good in both approaches up to $\sqrt{s}$ = 200 GeV. We note
in passing that the inclusion of PYTHIA (v 6.1) in UrQMD 2.0 and UrQMD 2.1
is essential in order to achieve a good description of the
data since the contribution from jet fragmentation is essential at
RHIC energies. Without jet production and fragmentation the UrQMD
model (version 1.3) underestimates the high momentum spectra
severely (similar to RQMD \cite{Sorge,Gyulassy}).

We continue with $pA$ reactions and show in the left part of Fig.
\ref{Fig_tpa} the results for the inverse slope parameters $T$ in
(\ref{slope}) for $\pi^\pm, K^+$ and $K^-$ mesons at midrapidity from
$pA$ reactions at 14.6 GeV$/c$ ($A$=Be, Al, Cu, Au), whereas the right
part includes the results at 450 GeV$/c$ ($A$=Be, S, Pb). The inverse
slope parameters from HSD are shown as open symbols connected by
solid lines and from UrQMD 1.3 (2.0) by open symbols (with crosses)
connected by dashed lines.  The full symbols (l.h.s.) correspond to the
midrapidity data ($\langle y_{lab}\rangle = 1.5, 1.7, 1.9$) from the E802
Collaboration \cite{E802_pA} while the NA44 data \cite{NA44_pA}
at $2.4\le y_{lab}\le 3.5, \ p_T\le 0.84$ GeV$/c$ for $K^+, K^-$ and
at $2.4\le y_{lab}\le 3.0, \ p_T =0.3\div 1.2$ GeV$/c$ (full diamonds)
and $3.1\le y_{lab}\le 4.0, \ p_T\le 0.64$ GeV$/c$ (full squares)
for $\pi^+$ are given on the right-hand side.

The overall reproduction of the transverse slope parameters at AGS and
SPS energies for targets from Be to Pb by both transport models is
sufficiently accurate. The inverse slope parameters from UrQMD 1.3 and HSD
differ for pions and antikaons by about 10\% at AGS energies.  These
differences can be traced back to the different production channels in
the transport models at lower energies:  in HSD the mesons are
essentially produced by string decays, which are strongly forward
peaked in momentum, whereas in UrQMD 1.3 they partly stem
from high mass resonance decays with an isotropic decay
in momentum space. At SPS energies the meson production
in both models proceeds via string formation and decay and leads to
very similar results. This also holds for RHIC energies where
additionally jet production and fragmentation contributes with
increasing bombarding energy.

We note in passing that when including the initial state 'Cronin
effect' (cf. Section VI) the inverse slope parameters $T$ in Fig.
\ref{Fig_tpa} increase by less than 5\% even for the Pb-target at 450
GeV/c. At 14.6 GeV/$c$ there is no effect (within statistics) for all
targets from Be to Au.

%---------------------------------------------------------------
\subsection{A+A reactions}

We continue with nucleus-nucleus collisions and recall that the
experimental $m_T$ spectra (at midrapidity)  from central C+C and Si+Si
collisions at 160 $A\cdot$GeV \cite{NA49_CCSi} are well described by
HSD and UrQMD 1.3.  This holds for the $\pi^-$, $K^+$ as well as $K^-$
spectra (cf. Fig. 1 in Ref. \cite{MT-prl}).

\begin{figure*}[!]
\centerline{\psfig{figure=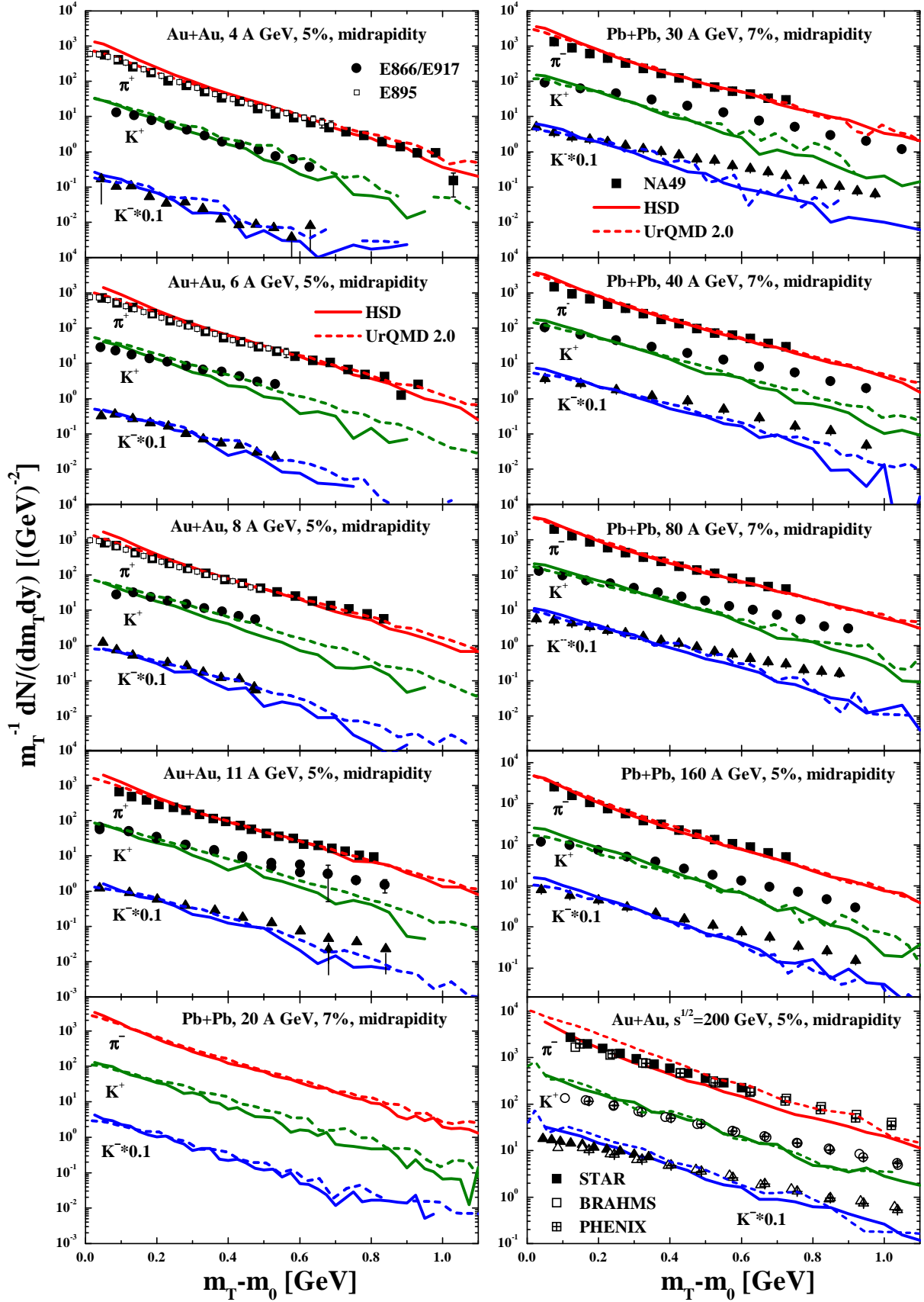,width=16cm}}
\caption{(Color online) Comparison of transverse mass spectra at
midrapidity from HSD (solid lines) and UrQMD 2.0 (dashed lines) for
$\pi^\pm, K^+$ and $K^- (\times 0.1)$ from central Au+Au (Pb+Pb)
reactions at 4, 6, 8, 11, 20 A$\cdot$GeV (left part) and at 30, 40, 80,
160 A$\cdot$GeV and $\sqrt{s}=200$ GeV (right part) with the data (full
and open symbols) from Refs.
\protect\cite{NA49_T,E866E917,E895,STAR,BRAHMS,PHENIX}. }
\label{Fig_mtaa}
\end{figure*}

This situation changes for central collisions of heavy nuclei like
Au+Au
or Pb+Pb. In Fig. \ref{Fig_mtaa} we display the calculated transverse mass
spectra at midrapidity from HSD (solid lines) and UrQMD 2.0 (dashed lines)
for $\pi^\pm, K^+$ and $K^- (\times 0.1)$ from central Au+Au (Pb+Pb)
reactions at 4, 6, 8, 11, 20 A$\cdot$GeV (left part) and at 30, 40, 80,
160 A$\cdot$GeV and $\sqrt{s}=200$ GeV (right part) with the data (full
and open symbols) from Refs.
\cite{NA49_T,E866E917,E895,STAR,BRAHMS,PHENIX} \footnote{Note that all
data from the NA49 Collaboration at 30 A$\cdot$GeV have to be
considered as 'preliminary'}.  At the lowest energy of 4 $A\cdot$GeV
the agreement between the transport approaches and the data is still
acceptable, however, severe deviations are visible in the $K^\pm$
spectra at top AGS, SPS and RHIC energies. Note that the $\pi^{\pm}$
spectra are reasonably described at all energies while the inverse
slope $T$ of the $K^\pm$ transverse mass spectra is underestimated
severely by about the same amount in both transport approaches (within
statistics).  The increase of the inverse $K^\pm$ slopes in heavy-ion
collisions with respect to $pp$ collisions, which is generated by
rescatterings of produced hadrons in the transport models, is only
small because the elastic meson-baryon scattering is strongly forward
peaked (in HSD and UrQMD 2.0) and therefore gives little additional
transverse momentum especially at midrapidity.

%---------------------------------------------------------------
\section{Testing alternative scenarios}

In this Section we will investigate different hadron/string scenarios,
which might contribute to the increase of the inverse slope parameter
of the $K^\pm$ mesons in central Au+Au collisions and are not included
in the 'default' versions of the transport approaches.

\subsection{In-medium potentials}

The question comes up whether the discrepancies shown in Fig.
\ref{Fig_mtaa} might be due to conventional hadronic medium effects.  In
fact, the $m_T$ slopes of kaons and antikaons at SIS energies (1.5 to
2 $A\cdot$GeV) were found to differ significantly \cite{KaoS}. As argued
in  \cite{Cass99} the different slopes could be traced back to
repulsive potentials of kaons with nucleons, which lead to a hardening
of the $K^+$ spectra,  and attractive antikaon-nucleon potentials,
which lead to a softening of the $K^-$ spectra. However,
the effect of such potentials was calculated within HSD and was found
to be of minor importance at AGS and SPS energies \cite{Cass99} since
the meson densities are comparable to or even larger than the baryon
densities at AGS energies and above.

Additional self energy contributions stem from $K^\pm$ interactions
with mesons which become increasingly important for higher bombarding
energies; however, $s$-wave kaon-pion interactions are weak due to
chiral symmetry arguments and $p$-wave interactions such as $\pi+K
\leftrightarrow K^*$ transitions are  suppressed substantially by the
approximately 'thermal' pion spectrum. A recent study on the kaon
potentials in hot pion matter gives kaon mass shifts of about $-52$ MeV
and vector potentials of $\sim +49$ MeV \cite{Fuchs} for a pion gas at
temperature $T$= 170 MeV.  We have employed even slightly larger
$K^\pm$ potentials in dynamical HSD calculations and achieved a
hardening of the $K^\pm$ spectra by less than 10\%.

\subsection{Overlapping strings}

Furthermore, we have pursued the idea of Refs. \cite{Sorge,Soff}
that the $K^\pm$ spectra could be hardened by string-string
interactions, which increase the effective string tension $\sigma$ and
thus the probability to produce mesons at high $m_T$
\cite{Soff,Ko_AMPT}. A related suggestion has been put forward in
\cite{carlos10}.

In order to estimate the largest possible effect of string-string
interactions we have assumed that for two overlapping strings the
string tension $\sigma$ is increased by a factor of two, for three
overlapping strings by a factor of three etc.  Here the overlap of
strings is defined geometrically assuming a transverse string radius
$R_s$, which according to the studies in Ref.  \cite{Geiss99} should be
$R_s \leq$ 0.25 fm. Additionally an overlap of the strings in
longitudinal direction is required which, however, does not involve any
new parameter since the dynamics of the string ends is directly
available from the HSD transport calculations. Based on these
assumptions (and $R_s$=0.25 fm as in Refs. \cite{carlos10}), we find
only a small increase of the inverse slope parameters at AGS energies,
where the string densities are low (dashed lines in Fig.
\ref{Fig_scenar} in comparison to the 'default' calculations (solid
lines)).  At SPS and RHIC energies the model gives a hardening of the
spectra by about 15\%  which, however, is still too low in comparison
to the data in Fig. \ref{Fig_mtaa}.  Despite of the large number of
'initial' strings in central Au+Au collisions at SPS or RHIC energies,
their average 'eigenvolume' (in the rest frame at break up) is low
($\approx 0.3$ fm$^3$) compared to the eigenvolume of a nucleon
($\approx 2$ fm$^3$) due to the fast string fragmentation process.
Consequently the critical density of color neutral strings may be about
7 times higher than the critical density of nucleons with respect to a
phase transition induced by percolation \cite{CCarlos}.

\subsection{Parton elastic scattering}

\begin{figure}[t]
\centerline{\psfig{figure=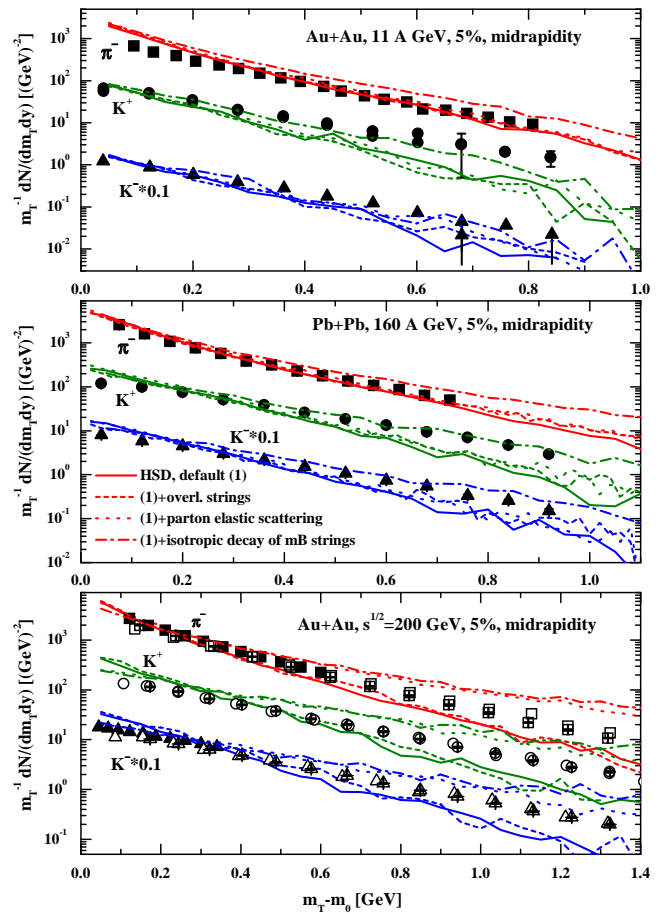,width=8.5cm}}
\caption{(Color online) The transverse mass spectra at midrapidity for
$\pi^-, K^+$ and $K^- (\times 0.1)$ from 5\% central Au+Au (Pb+Pb)
reactions at 11 A$\cdot$GeV, 160 A$\cdot$GeV and $\sqrt{s}=200$ GeV
calculated for different scenarios in HSD: the solid lines correspond
to the default calculations, the dashed lines to simulations with
overlapping strings, the dotted lines reflect calculations including
'parton elastic scattering' and the dot-dashed lines arise from
calculations with an isotropic decay of meson-baryon ($mB$) strings
(see text).  The experimental data are the same as in Fig.
\protect\ref{Fig_mtaa}.}
\label{Fig_scenar}
\end{figure}

We have examined another possibility which we denote as 'parton elastic
scattering'. To this end we have included elastic scattering of
unformed hadrons in the initial phase of the reaction -- before they
actually evolve to their full hadronic size -- employing the elastic
pion-nucleon cross section per quark degree of freedom. The results of
these simulations are shown in Fig. \ref{Fig_scenar} by the dotted
lines which give a much better reproduction of the experimental spectra
at $\sqrt{s}$ = 200 GeV, that are even slightly overestimated.  The
drawback on the other side is that the hadron rapidity spectra now are
overestimated by $\sim$ 20\% relative to the default calculation (and
the experimental data).  The increase in the slope parameter especially
for $K^\pm$ spectra at AGS and SPS energies, however, is only moderate
and a significant discrepancy remains with respect to the experimental
spectra.

\subsection{Isotropic decay of strings from meson-baryon collisions}

\begin{figure}[t]
\centerline{\psfig{figure=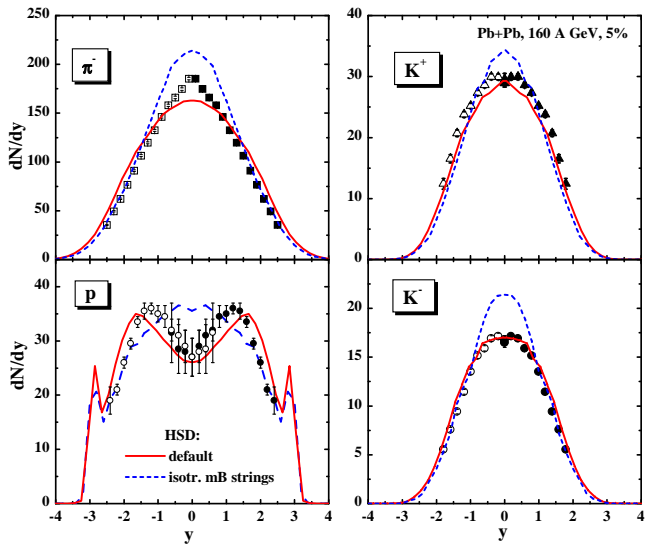,width=8.5cm}}
\caption{(Color online) The rapidity distribution for $\pi^-, K^+, K^-$
and protons from 5\% central Pb+Pb reactions at 160 A$\cdot$GeV.  The
solid lines correspond to the default HSD calculations, whereas the
dashed lines show the calculations with an isotropic decay of strings
from $mB$ collisions (see text).  The experimental data are taken from
Refs. \protect\cite{NA49_new,NA49pNew}.}
\label{Fig_y160iso}
\end{figure}

Following a further idea of Ref. \cite{Bleich99_mB} we explore the
possibility that the fragmentation of strings from $mB$ collisions
modifies in the dense and hot medium.  An extreme modification amounts
to assume an isotropic decay of $mB$ strings -- as employed in the RQMD
model \cite{Sorge} -- which leads to a reduction of longitudinal hadron
momenta and an enhancement of their transverse momenta. The results
of these simulations are displayed in Fig. \ref{Fig_scenar} for the
transverse mass spectra of pions, kaons and antikaons (dot-dashed
lines).  Apparently, this modified fragmentation function leads to much
better results in comparison to experiment (cf.  dot-dashed lines
in Fig.~\ref{Fig_scenar}).

The drawback with this scenario, however, is that the approximate
agreement with the longitudinal rapidity distributions for $\pi^-, K^+,
K^-$ and protons is destroyed as demonstrated in Fig.
\ref{Fig_y160iso} for 5\% central Pb+Pb reactions at 160 A$\cdot$GeV.
Here the solid lines correspond to the default HSD calculations whereas the
dashed lines stand for the calculations with an isotropic decay of $mB$
strings. Obviously, the latter scenario leads to an enhanced stopping
of the nuclei and larger meson production, which is not supported very
much by the data.  We admit, that a slightly more isotropic decay of
the $mB$ strings (with respect to the default calculations) would be
favored by the experimental data; however, we do not aim at fitting
data here. We mention - without explicit representation - that the
comparison shown in Fig.  \ref{Fig_y160iso} for central collisions at
160 A$\cdot$GeV holds qualitatively also at AGS and RHIC energies.

\subsection{High mass baryon resonances (UrQMD 2.1)}

\begin{figure}[t]
\centerline{\psfig{figure=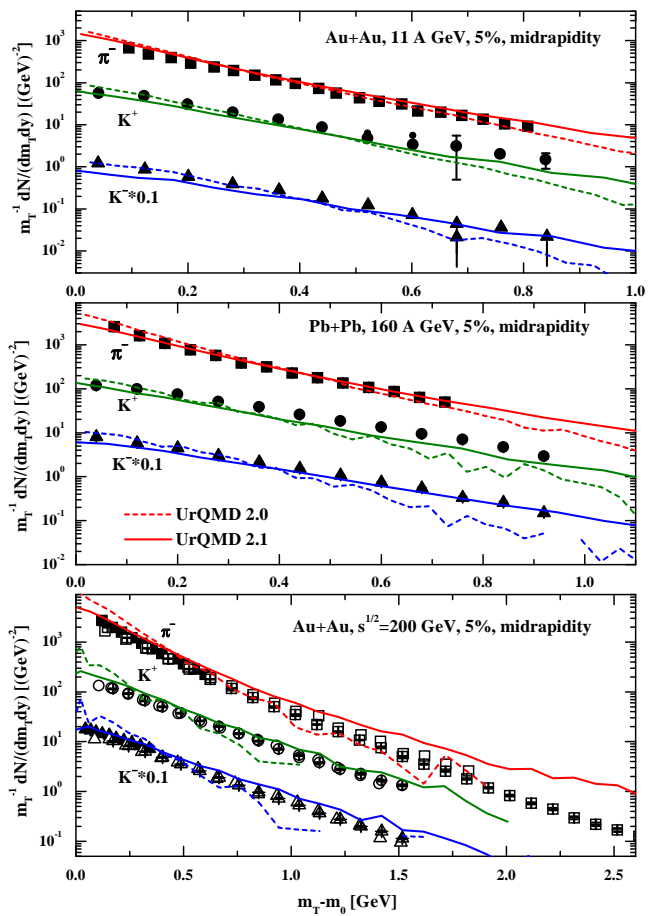,width=8.5cm}}
\caption{(Color online) The transverse mass spectra at midrapidity for
$\pi^-, K^+$ and $K^- (\times 0.1)$ from 5\% central Au+Au (Pb+Pb)
reactions at 11 A$\cdot$GeV, 160 A$\cdot$GeV and $\sqrt{s}=200$ GeV
calculated for different scenarios in UrQMD: the dashed lines
correspond to the default calculations (UrQMD 2.0), the solid lines to
simulations with high mass baryon resonances (UrQMD 2.1).  The
experimental data are the same as in Fig. \protect\ref{Fig_mtaa}.}
\label{Fig_mtU21}
\end{figure}

A further possibility to harden the meson and nucleon spectra in a
coupled-channel transport approach is to introduce higher mass baryon
resonances with masses $>2$ GeV as in RQMD \cite{Sorge}, that decay
isotropically and dominantly couple to the $\pi N$ channel and not to
multipion + nucleon channels. As described in Section II, the version
UrQMD 2.1 has been set up to incorporate this possibility and to
explore its consequences.  Since the 'massive' resonances decay
isotropically in their rest frame, this leads to enhanced transverse
flow as found in the previous Subsection.  The penalty of enhanced
hadron production (cf. Fig. \ref{Fig_y160iso}) is cured in UrQMD 2.1
since the high mass resonance decays produce 25\% less particles as
compared to a string of the same invariant mass.  In this way one can
produce larger hadron transverse momenta in the transport calculations
without increasing the meson multiplicities on average. Such a
'prescription' fits the data in the  AGS and SPS energy regime very
well (similar to RQMD \cite{Sorge,Hecke}).  However, the open question
remains, if such 'states' could be identified in meson-baryon
scattering and, furthermore, if they would survive in a dense hadronic
medium.

\begin{figure}[t]
\centerline{\psfig{figure=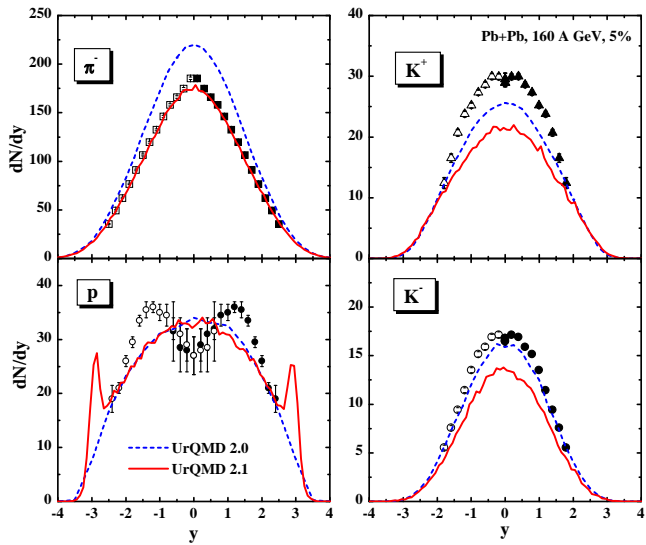,width=8.5cm}}
\caption{(Color online) The rapidity distribution for $\pi^-, K^+, K^-$
and protons from 5\% central Pb+Pb reactions at 160 A$\cdot$GeV.  The
dashed lines correspond to the default calculations from UrQMD 2.0,
whereas the solid lines show the calculations with high mass baryon
resonances (UrQMD 2.1).  The experimental data are taken from Refs.
\protect\cite{NA49_new,NA49pNew}.}
\label{Fig_y160u21}
\end{figure}

In Fig. \ref{Fig_mtU21}  we show the transverse mass spectra at
midrapidity for $\pi^-, K^+$ and $K^- (\times 0.1)$ from 5\% central
Au+Au (Pb+Pb) reactions at 11 A$\cdot$GeV, 160 A$\cdot$GeV and
$\sqrt{s}=200$ GeV calculated for different scenarios in UrQMD: the
dashed lines correspond to the default calculations (UrQMD 2.0), the
solid lines to simulations with high mass baryon resonances (UrQMD
2.1). Indeed, the description of the pion and $K^\pm$ improves
substantially in UrQMD 2.1 relative to UrQMD 2.0 (and HSD) at AGS and
SPS energies.  However, the penalty here is that now the spectra at the
top RHIC energy are overestimated at high transverse mass  for pions
and antikaons.

Moreover, the improvement in the spectral slope for pions and kaons at
AGS and SPS energies in UrQMD 2.1 does not go along with an improvement
in the rapidity spectra as shown in Fig. \ref{Fig_y160u21}. Though the
$\pi^-$ rapidity distribution can be well described for  5\% central Pb+Pb
reactions at 160 A$\cdot$GeV due to the reduction of pion production in
the decay scheme of the high mass states, the rapidity spectra of
$K^\pm$ mesons are also lower in UrQMD 2.1 (solid lines) compared to
UrQMD 2.0 (dashed lines) and the data. The proton stopping at
midrapidity is not much affected by the introduction of the high mass
states since they do not couple to the baryon-baryon channel.

We conclude, that the recipe of high mass states -- as incorporated in
UrQMD similar to RQMD -- does not solve the problem with the
$K^+/\pi^+$ ratio.

%**********************************************************************
\section{Cronin enhancement in $pA$ and $AA$ collisions}

The Cronin effect dates back to measurements of transverse hadron
spectra from $pA$ collisions at the SPS and ISR \cite{Cronin1,Cronin2}
and relates to a (moderate) enhancement of transverse momentum spectra
with increasing target mass $A$. Though the microscopic  origin of this
effect is not fully understood, most current interpretations are based
on semi-hard initial state gluon radiation of the propagating quarks in
the medium prior to a hard interaction vertex (cf. e.g. \cite{Kopel2}).
In this way the initial transverse momentum distribution of partons is
extended to higher average $\langle k_T^2\rangle$.  We discard a
further discussion of its physical origin and proceed with effective
simulations following Ref. \cite{Cass_Cron03}. In the latter work it
was suggested that such an enhancement of the intrinsic quark
transverse momentum spread $\langle k_T^2\rangle$ might be simulated in
the transport approach by increasing the average transverse momentum
spread of the quarks $\langle k_T^2\rangle$ with the number of previous
collisions of primary nucleons $N_{prev}$ as
\begin{equation}
\label{kt}
\langle k_T^2\rangle = \langle k_T^2\rangle_{pp} (1+\alpha N_{prev}).
\end{equation}
The parameter $\alpha \approx 0.4$ has been determined in Ref.
\cite{Cass_Cron03} by comparing to the experimental data for d+Au
collisions \cite{PHENIX_dAu,STAR_dAu} at $\sqrt{s}=200$ GeV. This
assumption is in line with an independent analysis within the Glauber
model by Papp et al. \cite{Papp}. Note, that $N_{prev}$ is not a
further parameter of the model but calculated dynamically in the
transport approach.

Whereas the different alternative scenarios discussed in  Section V
relate to various model assumptions, which are not directly controlled
by experimental data, the initial state Cronin enhancement is an
established experimental fact.  Consequently, it should be incorporated
in 'future' transport approaches (as in Ref.  \cite{Cass_Cron03}).  For
further details we refer the reader to the latter publication.

\subsection{$A+A$ reactions with initial state Cronin enhancement}

\begin{figure}[t]
\centerline{\psfig{figure=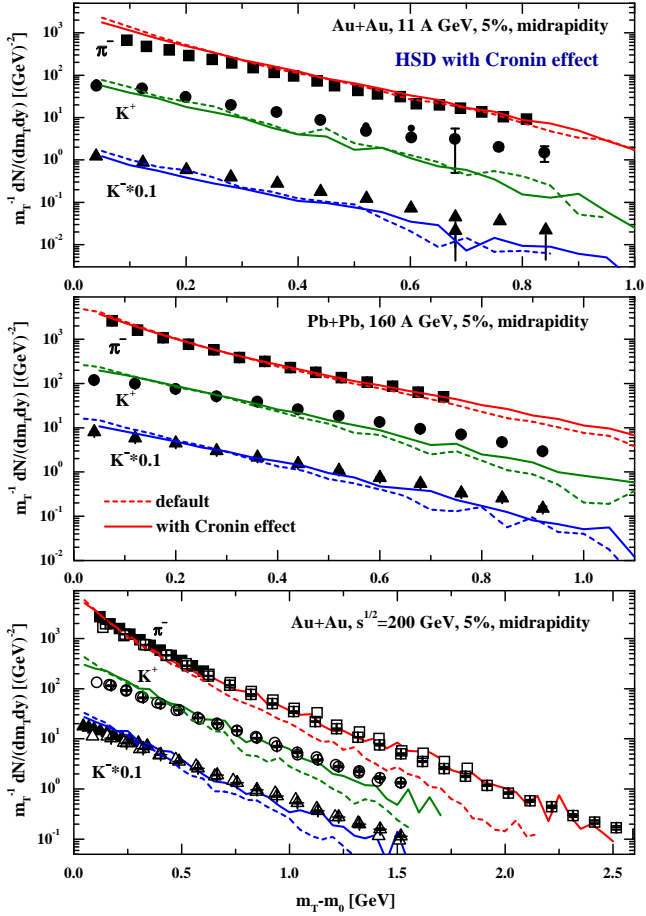,width=8.5cm}}
\caption{(Color online) The transverse mass spectra at midrapidity for
$\pi^-, K^+$ and $K^- (\times 0.1)$ from 5\% central Au+Au (Pb+Pb)
reactions at 11, 160 A$\cdot$GeV and $\sqrt{s}=200$ GeV calculated
within HSD including the Cronin effect (\ref{kt}) (solid lines) in
comparison to the default calculations (dashed lines).  The
experimental data are the same as in Fig. \protect\ref{Fig_mtaa}.}
\label{Fig_cronaa}
\end{figure}

Our results for the transverse mass spectra at midrapidity for $\pi^-,
K^+$ and $K^- (\times 0.1)$ from 5\% central Au+Au (Pb+Pb) reactions at
11 A$\cdot$GeV, 160 A$\cdot$GeV and $\sqrt{s}=200$ GeV as calculated
within HSD including the Cronin effect (\ref{kt}) (solid lines) are
presented in Fig. \ref{Fig_cronaa}  in comparison to the default
calculations (dashed lines).  The experimental data are the same as in
Fig.  \ref{Fig_mtaa}.  The description of the spectra becomes rather
good at RHIC energies, improves substantially at the SPS energy of 160
A$\cdot$GeV, but does not give any significant change at 11
A$\cdot$GeV. Consequently, the 'prehadronic' Cronin effect - as
implemented via (\ref{kt}) - is not responsible for the dramatic
increase of the $K^\pm$ slopes at AGS energies. Our calculations,
however, show that a hardening of the transverse mass spectra should
not only be attributed to the presence of collective flow, but also to
initial state semi-hard scattering of fast quarks in a dense medium;
the latter effect becomes more important with increasing $\sqrt{s}$. A
more detailed and quantitative analysis on the competition of initial
state Cronin enhancement and collective flow from pressure gradients at
SPS and RHIC energies we postpone to an upcoming study.

%------------------------------------------------------------------
\section{Inverse slope parameters and $\langle p_T\rangle$}

\begin{figure*}[!]
\centerline{\psfig{figure=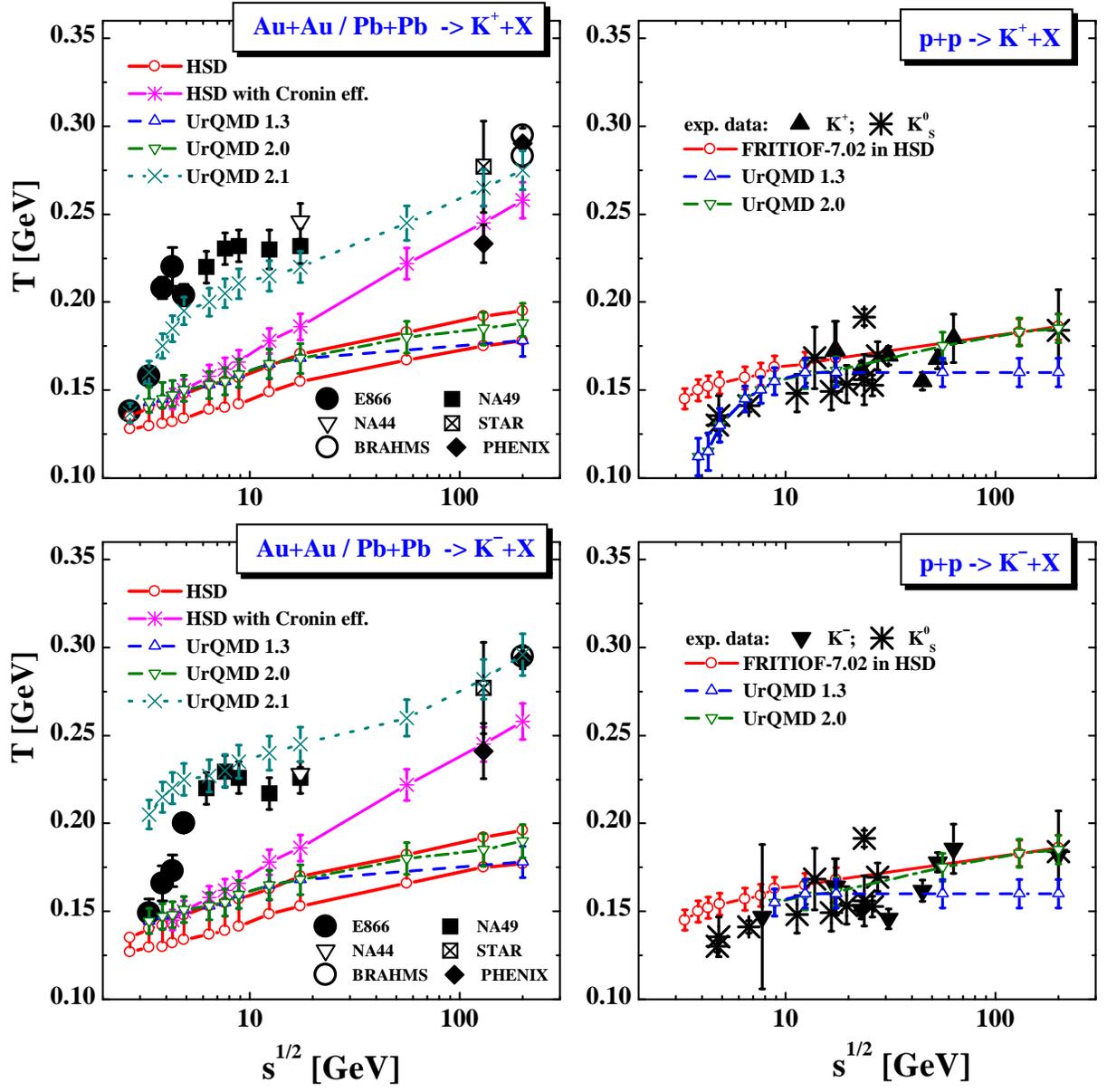,width=16cm}}
\caption{(Color online) Comparison of the inverse slope parameters $T$
for $K^+$ and $K^-$ mesons from central Au+Au (Pb+Pb) collisions (l.h.s.)
and $pp$ reactions (r.h.s.) at midrapidity as a function of the invariant
energy $\sqrt{s}$ from HSD (upper and lower solid lines with open circles),
UrQMD 1.3 (dashed lines with open upper triangles),
UrQMD 2.0 (dot-dashed lines with open lower triangles),
UrQMD 2.1 (dotted lines with crosses) with the data from Refs.
\protect\cite{E866E917,NA49_T,NA49_QM04,NA44,STAR,BRAHMS,PHENIX} for $AA$ and
\protect\cite{NA49_CCSi,Gazdz_pp,STAR} for $pp$ collisions
The upper and lower solid lines in the left diagrams result from
different limits of the HSD calculations as discussed in the text while
the solid lines with stars correspond to HSD calculations with
the Cronin initial state enhancement (\protect\ref{kt}).}
\label{Fig_Tcron}
\end{figure*}

\begin{figure*}[!]
\centerline{\psfig{figure=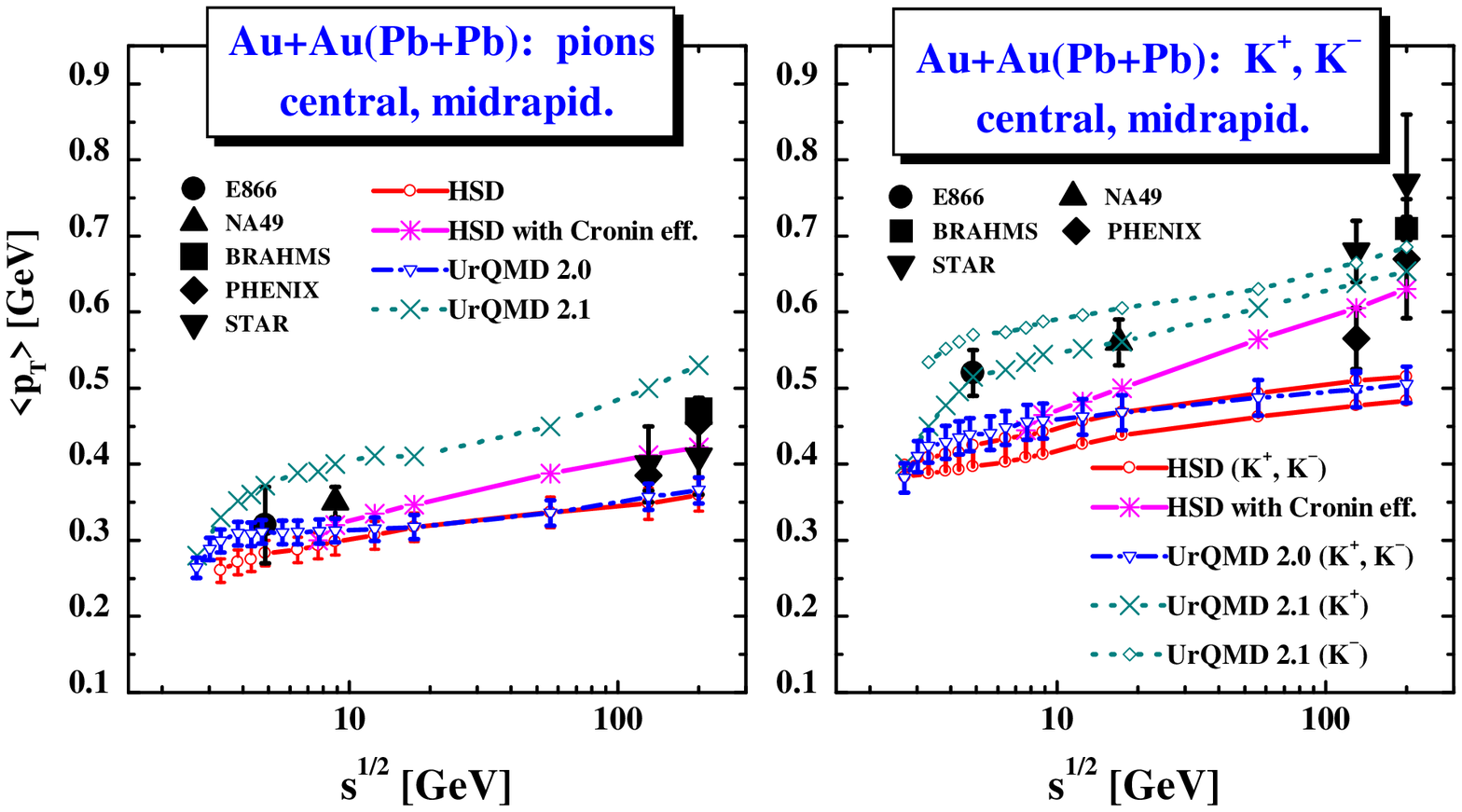,width=16cm}}
\caption{(Color online) The excitation function for the average values of
$p_T$ at midrapidity  for pions (left) and kaons (right) calculated
within HSD (upper and lower solid lines with open circles) and
UrQMD 2.0 (dot-dashed lines with open lower triangles),
UrQMD 2.1 (dotted lines with crosses)
in comparison to the data from Refs.
\protect\cite{E866_QM96,NA49_pt99,BRAHMS,PHENIX,STAR}.
The solid lines with stars correspond to the HSD calculations
with the Cronin effect (\protect\ref{kt}).
The dotted line with the open diamonds (right part)
gives $\langle p_T\rangle$ for $K^-$ from UrQMD 2.1.}
\label{Fig_pt}
\end{figure*}

Our findings are summarized in Fig. \ref{Fig_Tcron}, where the
dependence of the inverse slope parameter $T$ (defined by
Eq.~(\ref{slope})) on $\sqrt{s}$ is shown and compared to experimental
data (partly preliminary) from Refs.
\protect\cite{E866E917,NA49_T,NA49_QM04,NA44,STAR,BRAHMS,PHENIX} for central
Au+Au (Pb+Pb) collisions (l.h.s.) and \cite{NA49_CCSi,Gazdz_pp,STAR}
for $pp$ collisions (r.h.s.).  The upper and lower solid lines (with
open circles) on the l.h.s. in Fig. \ref{Fig_Tcron} correspond to
results from HSD calculations, where the upper and lower limits are due
to fitting the slope $T$ itself, an uncertainty in the repulsive
$K^\pm$-pion potential or the possible effect of string overlaps.  The
solid lines with stars correspond to the HSD calculations with the
Cronin effect  (\protect\ref{kt}).  The dashed lines with open upper
triangles represent slope parameters from UrQMD 1.3, the dot-dashed
lines with open lower triangles correspond to UrQMD 2.0 results, which
are well within the limits obtained from the different HSD calculations
without the Cronin enhancement.  The dotted lines with crosses show the
UrQMD 2.1 results that incorporate the high mass resonance states.

The slope parameters from $pp$ collisions (r.h.s. in Fig.
\ref{Fig_Tcron}) are seen to increase smoothly with energy both in the
experiment (full symbols) and in the HSD calculations (full lines with
open circles). The error bars related to our calculations are due to an
uncertainty in extracting the slope parameter from different $m_T$
intervals. The UrQMD 1.3 results are shown as open upper triangles
connected by the solid line and systematically lower than the slopes
from HSD at all energies. When including jet production and
fragmentation via PYTHIA in UrQMD 2.0 (dot-dashed lines with open lower
triangles) the results become similar to HSD above
$\sqrt{s}$ = 10 GeV demonstrating the importance of jets in $pp$
reactions at high energy.

Coming back to the slope parameters of $K^\pm$ mesons for central
Au+Au/Pb+Pb collisions (l.h.s. of Fig.  \ref{Fig_Tcron}) we find that
the Cronin initial state enhancement indeed improves the description of
the data at RHIC energies, however, does not give any sizeable
enhancement at AGS energies.  Here UrQMD 2.1 (dashed lines with
crosses) with the high mass resonance states performs better for $K^+$
mesons, but overestimates the $K^-$ slopes at AGS as well as upper SPS
energies.

An alternative way of representing the data is provided by the
excitation function for the average transverse momentum
$\langle p_T\rangle$ for pions (l.h.s.) and kaons (r.h.s.)
in Fig. \ref{Fig_pt} for the different transport models in comparison
to the data available.
HSD: upper and lower solid lines with open circles,
UrQMD 2.0: dot-dashed lines with open lower triangles and
UrQMD 2.1: dotted lines with crosses; the data are from
Refs. \cite{E866_QM96,NA49_pt99,BRAHMS,PHENIX,STAR}.
The dotted line with the open diamonds (r.h.s.) shows $\langle
p_T\rangle$ for $K^-$ from UrQMD 2.1. Furthermore, the solid lines with
stars correspond to the HSD calculations with the Cronin effect, which
gives a good description of the average $\langle p_T\rangle$ in case of
pions and significantly improves the average $\langle p_T\rangle$ for
kaons as a function of $\sqrt{s}$ (with respect to default HSD and
UrQMD 2.0 calculations).  The UrQMD 2.1 calculations overestimate
$\langle p_T\rangle$ for pions at all energies by 10 to 15 \%  as well
as for $K^-$ at AGS and SPS energies in line with the trends from Fig.
\ref{Fig_Tcron}; however, the average $p_T$ for $K^+$ mesons is rather
well described.

\section{Thermodynamics in the $T-\mu_B$ plane}

\begin{figure*}[!]
\centerline{\psfig{figure=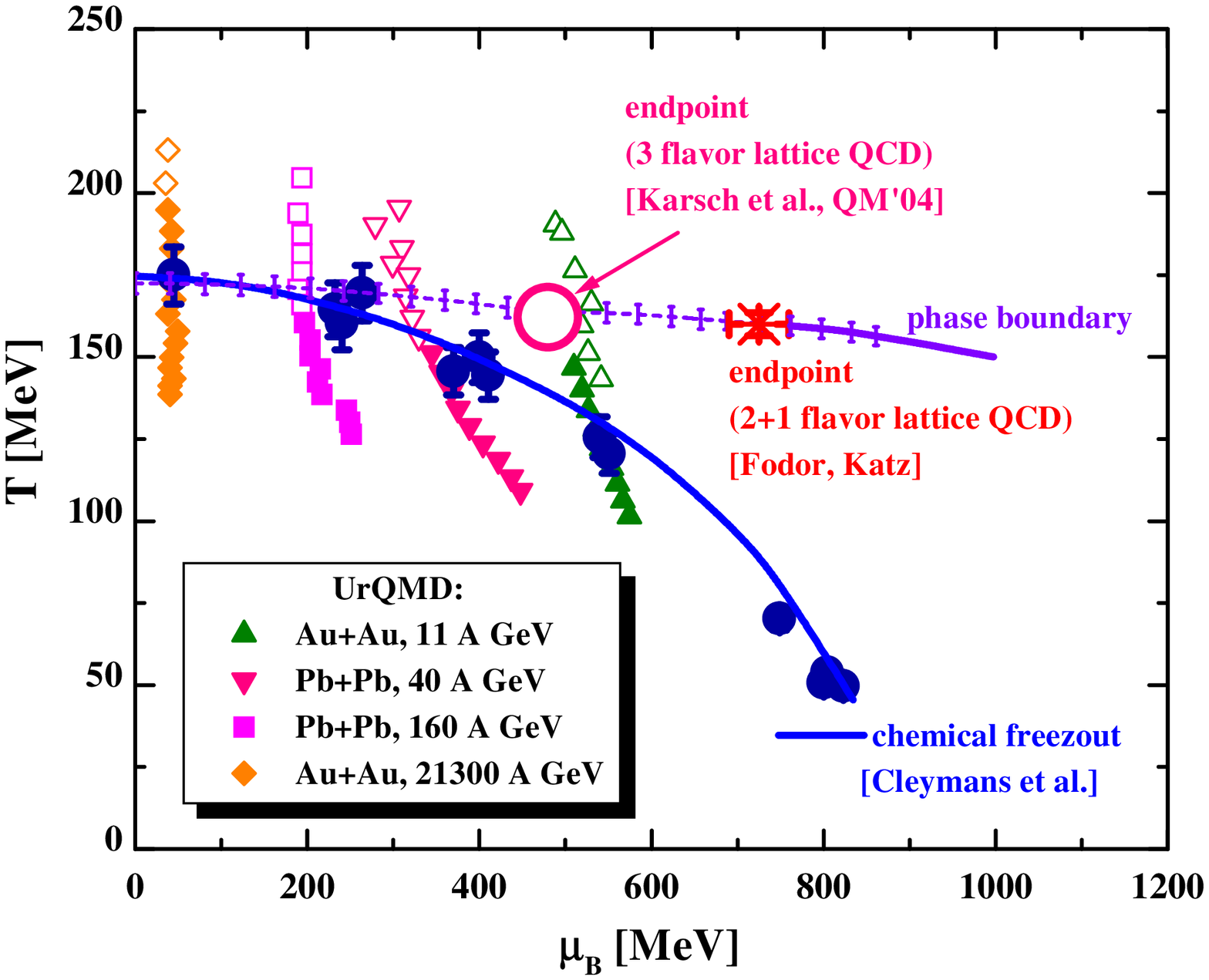,width=13.5cm}}
\caption{(Color online)  Schematic phase diagram in the $T-\mu_B$
plane.  The solid line characterizes the universal chemical freeze-out
line from Cleymans et al. \protect\cite{Cleymans} whereas the full dots
(with errorbars) denote the 'experimental' chemical freeze-out
parameters from Ref.  \protect\cite{Cleymans}. The various symbols
represent temperatures $T$ and chemical potentials $\mu_B$ extracted
from UrQMD 1.3 transport calculations in central Au+Au (Pb+Pb)
collisions at 21.3 A$\cdot$TeV, 160, 40 and 11 A$\cdot$GeV
\protect\cite{Bravina} (see text). The large open circle and the star
indicate the tri-critical endpoints from lattice QCD calculations by
Karsch et al.  \protect\cite{Karsch2} and Fodor and Katz
\protect\cite{Fodor}, respectively. The horizontal line with errorbars
is the phase boundary from \protect\cite{Fodor}.}
\label{Fig_QCD}
\end{figure*}

This still leaves us with the question of the origin of the rapid
increase of the $K^\pm$ slopes with invariant energy for central Au+Au
collisions at AGS energies and the constant slope at SPS energies (the
'step'), which is  missed in all transport approaches presently
employed.  We recall that higher transverse particle momenta either
arise from repulsive self energies -- in mean-field dynamics -- or from
collisions, which reduce longitudinal momenta in favor of transverse
momenta \cite{HORST,CaMo}.  As shown above in Fig. \ref{Fig_Tcron}
conventional hadron self-energy effects and (pre-)hadronic and hadronic
binary collisions are insufficient to describe the dramatic increase of
the $K^\pm$ slopes as a function of $\sqrt{s}$ as well as the initial
state Cronin enhancement. Only when including high mass resonances in
UrQMD 2.1 -- with a particular decay pattern -- we may end up with
reasonable results for $K^+$ mesons, however, fail more severely for
pions as well as antikaons.  These failures of 10 to 15\% indicate an
additional mechanism for the generation of the pressure that is
observed experimentally especially at AGS energies.

We speculate -- following the previous study \cite{MT-prl} -- that
partonic degrees of freedom should be responsible for this effect
already at $\sim$ 5 $A\cdot$GeV. Our arguments here are based on a
comparison of the thermodynamic parameters $T$ and $\mu_B$ extracted
from the transport models in the central overlap regime of Au+Au
collisions \cite{Bravina} with the experimental systematics on chemical
freeze-out configurations \cite{Cleymans} in the $T,\mu_B$ plane. The
solid line in Fig.  \ref{Fig_QCD} characterizes the universal chemical
freeze-out line from Cleymans et al. \cite{Cleymans} whereas the full
dots with errorbars denote the 'experimental' chemical freeze-out
parameters - determined from the thermal model fits to the experimental
ratios - taken from Ref.  \cite{Cleymans}. The various symbols (in vertical
sequence) represent temperatures $T$ and chemical potentials $\mu_B$
extracted from UrQMD 1.3 transport calculations in central Au+Au
(Pb+Pb) collisions at 21.3 A$\cdot$TeV, 160, 40 and 11 A$\cdot$GeV
\cite{Bravina} as a function of the reaction time in the center-of-mass
(from top to bottom).
The open symbols denote nonequilibrium configurations and correspond to
$T$ parameters extracted from the transverse momentum distributions,
whereas the full symbols denote configurations in approximate pressure
equilibrium in longitudinal and transverse direction.

During the nonequilibrium phase (open symbols) the transport
calculations show much higher temperatures (or energy densities) than
the 'experimental' chemical freeze-out configurations at all bombarding
energies ($\geq$~11~A$\cdot$GeV).  These numbers are also higher than
the tri-critical endpoints extracted from lattice QCD calculations by
Karsch et al.  \cite{Karsch2} (large open circle) and Fodor and Katz
\cite{Fodor} (star with horizontal error bar). Though the QCD lattice
calculations differ substantially in the value of $\mu_B$ for the
critical endpoint, the critical temperature $T_c$ is closer to
160 MeV in both calculations, while the energy density is of the order
of 1 GeV/fm$^3$ or even below. Nevertheless, this diagram shows that at
RHIC energies one encounters more likely a cross-over between the
different phases when stepping down in temperature during the expansion
phase of the hot fireball.

This situation might change at lower SPS or AGS (as well as  new GSI
SIS-300 \cite{GSIprop}) energies, where for sufficiently large chemical
potentials $\mu_B$ the cross over should change to a first order
transition \cite{Shuryak}, i.e. beyond the tri-critical point in the
($T,\mu_B$) plane.  Nevertheless, Fig.  \ref{Fig_QCD} demonstrates that
the transport calculation show temperatures (energy densities) well
above the phase boundary (horizontal line with errorbars from Ref.
\cite{Fodor}) in the very early phase of the collisions, where hadronic
interactions yield only a pressure $p \sim 0.1 -- 0.14 \epsilon$ (with
$\epsilon$ denoting the energy density). We recall that for massless
partonic degrees of freedom we have $p \sim \epsilon/3$ and
consequently a much higher pressure. This argument is well in line with
the studies on elliptic flow at RHIC energies, that is underestimated
by $\sim$ 30\% at midrapidity in the HSD approach \cite{Brat03} and by
a factor of $\sim 2$ in the UrQMD 1.3 model \cite{Bleicher_v2}.  Only
strong pre-hadronic/partonic interactions might cure this problem.

As shown in Ref. \cite{Mihlos}, in order to describe the elliptic flow
seen experimentally at RHIC in a parton cascade model, one has to
 employ parton cross sections up to 45 mb.  However, such strong
interactions are incompatible with perturbative QCD, which gives cross
sections that are lower by more than an order of magnitude
\cite{Mihlos}. We speculate that such strong nonperturbative
interactions on the partonic level are responsible for the large
pressure generation in the very early phase of intermediate energy
nucleus-nucleus collisions.

%**********************************************************************
\section{Summary}

Summarizing this work, we point out that baryon stopping and hadron
production in central Au+Au (or Pb+Pb) collisions is quite well
described in the independent transport approaches HSD and UrQMD 2.0.
Also the 'longitudinal' rapidity distributions of protons, pions,
kaons, antikaons and hyperons are similar in both models and in
reasonable agreement with available data.  The exception are the pion
rapidity spectra at the highest AGS energy and lower SPS energies,
which are overestimated by both models \cite{Weber02}.  As a
consequence the HSD and UrQMD 2.0 transport approaches underestimate
the experimental maximum of the $K^+/\pi^+$ ratio  at $\sim$ 20 to 30
A$\cdot$GeV and UrQMD underestimates also the data at RHIC energies.
This also holds for the new version UrQMD 2.1 introduced here
additionally, which incorporates high mass resonance states above 2 GeV
with a particular decay pattern.   This recipe cures the problem of
the overestimated pion yield at AGS and SPS energies
but also reduces the $K^\pm$ yield and therefore has only little
effect on the $K^+/\pi^+$ ratio.

We have found that the inverse slope parameters $T$ for $K^\pm$ mesons
from the HSD and UrQMD 2.0 transport models are practically independent of
system size from $pp$ up to central Pb+Pb collisions and show only a
slight increase with collision energy. The calculated transverse mass
spectra are in reasonable agreement with the experimental results for
$pp$ reactions at all bombarding energies investigated as well as
central collisions of light nuclei (C+C and Si+Si) (cf. Ref.
\cite{MT-prl}). The rapid increase of the inverse slope parameters of
kaons for central collisions of heavy nuclei (Au+Au or Pb+Pb) found
experimentally in the AGS energy range, however, is not reproduced by
both models in their default version (see Fig.~\ref{Fig_Tcron}).

We have, furthermore, discussed a couple of scenarios in Section V
to improve the description of the experimental data.
However, none of the alternative scenarios leads to a fully satisfactory
reproduction of all the experimental data.
It is worth pointing out that a substantial hardening of the $K^\pm$
slope results from an initial state Cronin enhancement at higher SPS
and RHIC energies, that should not be attributed to collective flow (or
initial pressure). This Cronin enhancement -- as known experimentally
from $p+A$ reactions at SPS energies -- does not signal a phase
transition to a quark-gluon-plasma.

We thus have argued - in comparison to lattice QCD calculations  at
finite temperature and baryon chemical potential $\mu_B$ from Refs.
\cite{Fodor} and \cite{Karsch2} as well as the experimental systematics
in the chemical freeze-out parameters (cf. Fig. \ref{Fig_QCD}) -- that
the missing pressure should be generated in the early phase of the
collision by nonperturbative partonic interactions because the strong
hadronic interactions  don't produce it in the later stages.

The interesting finding of our analysis is, that pre-hadronic/partonic
degrees of freedom seem to play a substantial role in central Au+Au
collisions already at AGS energies above $\sim$ 5 $A\cdot$GeV.  This
finding is in line with the 'observation' that  the excitation function
of the proton elliptic flow $v_2$ also indicates a softening of the
equation of state with increasing energy at 4-5 $A\cdot$GeV in Au+Au
collisions \cite{Dani,Sahu,Toneev}.  However, further detailed
experimental and theoretical studies also on collective flow properties
from nucleus-nucleus collisions will be necessary to unravel the
various aspects of relativistic many-body systems.

\section*{Acknowledgement}
The authors acknowledge inspiring discussions with K. Gallmeister,
C.~Greiner, C.~M.~Ko, Z. W. Lin, A. Peshier, K. Redlich, V. D. Toneev
and H. Weber.  Furthermore, they are indepted to M.~Ga\'zdzicki and
J. Klay for providing experimental data in numerical form.
E.L.B. is supported by Deutsche Forschungsgemeinschaft (DFG) and GSI.
S.A.B. acknowledges support  by RIKEN, Brookhaven National Laboratory
and DOE grants DE-FG02-96ER40945 and DE-AC02-98CH10886 as well as an
Outstanding Junior Investigator Award (DOE grant DE-FG02-03ER41239).
M.B. and M.R. are supported by GSI.

%-----------------------------------------------------------------------


\begin{thebibliography}{99}
\bibitem{QM02}
    {\it Quark Matter 2002}, Nucl. Phys. A {\bf 715}, 1 (2003).
\bibitem{Rafelski}
    J.~Rafelski and B.~M\"uller, Phys. Rev. Lett. {\bf 48}, 1066 (1982);
    P. Koch, B. M\"uller and J. Rafelski, Phys. Rep. {\bf 142} 167 (1986).
\bibitem{Karsch}
    F. Karsch {\it et al.}, Nucl. Phys. B {\bf 502}, 321 (2001).
\bibitem{HORST}
    H. St\"ocker and W. Greiner, Phys. Rep. {\bf 137}, 277 (1986).
\bibitem{exita}
    W. Cassing {\it et al.}, Nucl. Phys. A {\bf 674}, 249 (2000).
\bibitem{Weber98}
    H. Weber, C. Ernst, M. Bleicher {\it et al.},
    Phys. Lett. B {\bf 442}, 443 (1998).
\bibitem{Fodor}
    Z. Fodor and S. D. Katz,  JHEP {\bf 0203}, 014 (2002);
    Z. Fodor, S. D. Katz, and K. K. Szabo,
    Phys. Lett. B {\bf 568}, 73 (2003).
% hep-lat/0208078.
\bibitem{bjorken}
    J.D. Bjorken, Phys. Rev. D {\bf 27}, 140 (1983).
\bibitem{NA49_new}
     S. V.Afanasiev {\it et al.}, NA49 Collaboration,
     Phys. Rev. C {\bf 66}, 054902 (2002).
\bibitem{NA49_T}
    V. Friese {\it et al.}, NA49 Collaboration,
    J. Phys. G {\bf 30}, 119 (2004).
%   nucl-ex/0305017,
\bibitem{survey}
    J. L. Nagle and T. S. Ullrich, nucl-ex/0203007.
\bibitem{PHENIX_dAu}  % p+p, d+Au
    S.S. Adler {\it et al.}, PHENIX Collaboration,
    Phys. Rev. Lett. {\bf 91}, 072303 (2003);  %    nucl-ex/0306021.
    nucl-ex/0308006.
\bibitem{STAR_dAu} % d+Au, Au+Au
    J. Adams {\it et al.}, STAR Collab.,
    Phys. Rev. Lett. {\bf 91}, 072304 (2003); % nucl-ex/0306024
    {\it ibid.} {\bf 91}, 172302 (2003). % nucl-ex/0305015
\bibitem{BRAHMS_dAu} % d+Au, Au+Au
    I. Arsene {\it et al.},
    Phys. Rev. Lett. {\bf 91}, 072305 (2003); % nucl-ex/0307003.
\bibitem{Wang}
    X. N. Wang, Phys. Rev. C {\bf 58}, 2321 (1998).
\bibitem{Baier}
    R. Baier {\it et al.}, Ann. Rev. Nucl. Part. Sci. {\bf 50}, 37 (2000).
\bibitem{Goren}
    M. I. Gorenstein, M. Ga\'zdzicki, and K. Bugaev,
    Phys. Lett. B {\bf 567}, 175 (2003).
%   hep-ph/0303041.
\bibitem{SMES}
    M. Gazdzicki and M. I. Gorenstein,
    Acta Phys. Polon. B {\bf 30}, 2705 (1999).
\bibitem{Weber02}
    H. Weber, E. L. Bratkovskaya, W. Cassing, and H. St\"ocker,
     Phys. Rev. C {\bf 67}, 014904 (2003).
\bibitem{MT-prl}
    E. L. Bratkovskaya, S. Soff, H. St\"ocker, M. van Leeuwen, and
    W. Cassing,
    Phys. Rev. Lett., {\bf 92}, 032302 (2004).
% nucl-th/0307098,
\bibitem{UrQMD1}
    S.A.~Bass {\it et al.}, Prog. Part. Nucl. Phys. {\bf 42}, 255 (1998).
\bibitem{UrQMD2}
    M.~Bleicher {\it et al.},  J. Phys. G {\bf 25}, 1859 (1999).
\bibitem{Geiss}
    J. Geiss, W. Cassing, and C. Greiner, Nucl. Phys. A {\bf 644}, 107 (1998).
\bibitem{Cass99}
	W. Cassing and E. L. Bratkovskaya, Phys. Rep. {\bf 308}, 65 (1999).
\bibitem{Cass_Cron03}
    W. Cassing, K. Gallmeister, and C. Greiner,
     Nucl. Phys. A {\bf 735}, 277 (2004).
\bibitem{HSDcode}
	The HSD transport code is available from:
	http://www.th.physik.uni-frankfurt.de/$\sim$brat/hsd.html
\bibitem{BMJS}
    P. Braun-Munzinger, J. Stachel, J.P. Wessels, and N. Xu,
    Phys. Lett. B {\bf 344}, 43 (1995);
    {\it ibid.} {\bf 365},1 (1996);
    P. Braun-Munzinger, J. Cleymans, H. Oeschler and K. Redlich,
    Nucl. Phys. A {\bf 697}, 902 (2002).
\bibitem{Bravina}
    L. V. Bravina {\it et al.}, Phys. Rev. C {\bf 60}, 024904 (1999).
     Nucl. Phys. A {\bf 698}, 383 (2002).
\bibitem{Brat00}
     E. L. Bratkovskaya, W. Cassing, C. Greiner {\it et al.},
       Nucl. Phys. A {\bf 675}, 661 (2000).
\bibitem{Ko_AMPT}
     Z. W. Lin {\it et al.}, Nucl. Phys. A {\bf 698}, 375 (2002).
\bibitem{PDG}
    K.~Hagiwara {\it et al.}, (Review of Particle Properties),
    Phys. Rev. D {\bf 66}, 010001 (2002).
\bibitem{PYTHIA}
    H.-U. Bengtsson and T. Sj\"ostrand,
    Comp. Phys. Commun. {\bf 46}, 43 (1987).
\bibitem{Sorge}
    H. Sorge, Phys. Rev. C {\bf 52}, 3291 (1995).
\bibitem{LUND}
    B. Andersson {\it et al.}, Z. Phys. C {\bf 57}, 485 (1993).
\bibitem{Ehehalt}
    W. Ehehalt and W. Cassing, Nucl. Phys. A {\bf 602}, 449 (1996).
\bibitem{E866E917} % (pi+,K+,K- midrap., K+/pi+ ratio, K-/K+ ratio)
    L. Ahle {\it et al.}, E866 and E917 Collaboration,
    Phys. Lett. B {\bf 476}, 1 (2000);
    {\it ibid.} {\bf 490}, 53 (2000);
    Phys. Rev. C {\bf 58}, 3523 (1998).
\bibitem{E895} % new pion spectra
    J. L. Klay {\it et al.}, E895 Collaboration,
    Phys. Rev. C {\bf 68}, 054905 (2003).
%   nucl-ex/0306033,
\bibitem{E891Lam} % (Lambda's dN/dy at midrapidity, i.e. Lambdas yiled)
      S. Ahmad {\it et al.}, E891 Collaboration,
    Phys. Lett. B {\bf 382}, 35 (1996);
      C. Pinkenburg {\it et al.}, E866 Collaboration,
        Nucl. Phys. A {\bf 698}, 495c (2002).
\bibitem{NA49_Lam} % (Lambdas)
    A.~Mischke {\it et al.}, NA49 Collaboration,
    J. Phys. G. {\bf 28}, 1761 (2002);
     Nucl. Phys. A {\bf 715}, 453 (2993).
%    nucl-ex/0209002,
\bibitem{Antiori} % (Lambdas)
    F. Antinori {\it et al.}, WA97 Collaboration,
    Nucl. Phys. A {\bf 661}, 130c (1999).
\bibitem{NA49_QM04} % 20, 30 GeV data
    M. Gazdzicki {\it et al.}, NA49 Collaboration,
	nucl-ex/0403023.
\bibitem{BRAHMS}
    D. Ouerdane {\it et al.}, BRAHMS Collaboration,
    Nucl. Phys. A {\bf 715}, 478 (2003);
    J. H. Lee {\it et al.},  J. Phys. G {\bf 30}, S85 (2004).
\bibitem{PHENIX}
    S. S. Adler {\it et al.}, PHENIX Collaboration,
	Phys. Rev. C {\bf 69}, 024904 (2004); %    nucl-ex/0307010;
	nucl-ex/0307022.
\bibitem{STAR}
    C. Adler {\it et al.}, STAR Collaboration, nucl-ex/0206008;
    O. Barannikova {\it et al.}, Nucl. Phys. A {\bf 715}, 458 (2003);
%   nucl-ex/0210034;
    K. Filimonov {\it et al.}, hep-ex/0306056.
\bibitem{Andronic}
	A. Andronic and P. Braun-Munzinger,
	hep-ph/0402291.
\bibitem{FOPI} % pi-/pi+ ratio
    P. Crochet {\it et al.}, FOPI Collaboration,
    J. Phys. G {\bf 27}, 267 (2001).
\bibitem{MarcoSQM03}
    M. van Leeuwen {\it et al.}, NA49 Collaboration,
    Nucl. Phys. A {\bf 715}, 161 (2003).
\bibitem{Brat03}
    E. L. Bratkovskaya, W. Cassing and H. St\"ocker,
    Phys. Rev. C {\bf 67}, 054905 (2003).
\bibitem{Fischer}
     H. G. Fischer {\it et al.}, NA49 Collaboration,
    Nucl. Phys. A {\bf 715}, 118c (2003).
\bibitem{Blobel74}
    V. Blobel {\it et al.}, Nucl. Phys. B {\bf 69}, 454 (1974).
\bibitem{Gyulassy}
    V. Topor Pop, M. Gyulassy, J. Barrette, C. Gale, X.N. Wang,
    N. Xu, and K. Filimonov,
    Phys. Rev. C {\bf 68}, 054902 (2003).
%   nucl-th/0209089.
\bibitem{E802_pA}
    T. Abbott {\it et al.}, E802 Collaboration,
    Phys. Rev. D {\bf 45}, 3906 (1992).
\bibitem{NA44_pA} % pA at SPS
    H. Boggild {\it et al.}, NA44 Collaboration,
    Phys. Rev. C {\bf 59}, 328 (1999).
\bibitem{KaoS}
    A. F\"orster {\it et al.}, KaoS Collab.,
    J. Phys. G {\bf 28}, 2011 (2002).
\bibitem{Fuchs}
    B. V. Martemyanov {\it et al.}, nucl-th/0212064.
\bibitem{Soff} % (string enhancement)
      S. Soff {\it et al.}, Phys. Lett B {\bf 471}, 89 (1999).
\bibitem{carlos10}
    J. Dias de Deus, E. G. Ferreiro, C. Pajares, and R. Ugoccioni,
    Phys. Lett. B {\bf 581}, 156 (2004); %  hep-ph/0303220
    hep-ph/0304068.
\bibitem{Geiss99}
    J. Geiss, C. Greiner, E.L. Bratkovskaya {\it et al.},
    Phys. Lett. {\bf B447}, 31 (1999).
\bibitem{CCarlos}
     E.G. Ferreiro, C. Pajares, and D. Sousa,
    Phys. Lett. B {\bf 422}, 314 (1998);
     M.A. Braun and C. Pajares, Eur. Phys. J. C {\bf 16}, 349 (2000).
\bibitem{Bleich99_mB}
    M. Bleicher, C. Spieles, C. Ernst {\it et al.},
    Phys. Lett. B {\bf 447}, 227 (1999).
\bibitem{Hecke} % RQMD
    H. van Hecke {\it et al.}, Phys. Rev. Lett. {\bf 81}, 5764 (1998).
\bibitem{NA49pNew}
       C.E. Copper {\it et al.}, NA49 Collaboration,
    Nucl. Phys. A {\bf 661}, 362c (1999).
\bibitem{Cronin1}
    J.W. Cronin {\it et al.}, Phys. Rev. D {\bf 11}, 3105 (1975).
\bibitem{Cronin2}
    D. Antreasyan {\it et al.}, Phys. Rev. D {\bf 19}, 764 (1979).
\bibitem{Kopel2}
    B.Z. Kopeliovich, J. Nemchik, A. Sch\"afer and A.V. Tarasov,
    Phys. Rev. Lett. {\bf 88}, 232302 (2002).
\bibitem{Papp}
    G. Papp, P. Levai, and G. Fai, nucl-th/0307062.
\bibitem{NA44} % T at 160 GeV/A
    I.G. Bearden {\it et al.}, NA44 Collaboration,
    nucl-ex/0202019.
\bibitem{NA49_CCSi}
    I. Kraus {\it et al.}, NA49 Collaboration,
     J. Phys. G {\bf 30}, 5583 (2004).
%    nucl-ex/0306022,
\bibitem{Gazdz_pp}  % compilation on T-slope from pp
    M. Kliemant, B. Lungwitz, and M. Ga\'zdzicki,
    hep-ex/0308002.
\bibitem{E866_QM96} % <p_t>
        L. Ahle {\it et al.}, E866 Collaboration,
    Nucl. Phys. A {\bf 610}, 139c (1966).
\bibitem{NA49_pt99}  % <pt> at 160 GeV
    J. B\"achler {\it et al.}, NA49 Collaboration,
    Nucl. Phys. A {\bf 661}, 45c (1999).
\bibitem{CaMo}
    W. Cassing and U. Mosel,
    Prog. Part. Nucl. Phys. {\bf 25}, 235 (1990).
\bibitem{Cleymans}
    J. Cleymans and K. Redlich, Phys. Rev. C {\bf 60}, 054908 (1999).
\bibitem{Karsch2}
    F. Karsch,  talk given in {\it Quark Matter 2004},
    Oakland, January 11-17, 2004; http://qm2004.lbl.gov
%  hep-lat/0309116 - "old end point"
\bibitem{GSIprop}
    'An International Accelerator Facility for Beams of Ions and
    Anti\-protons';  http://www.gsi.de/GSI-Future/cdr/.
\bibitem{Shuryak}
    E. V. Shuryak, Nucl. Phys. A {\bf 661}, 119c (1999).
\bibitem{Bleicher_v2}
      M. Bleicher and H. St\"ocker, Phys. Lett. B {\bf 526}, 309 (2002).
\bibitem{Mihlos}
     D. Molnar and M. Gyulassy, Nucl. Phys. A {\bf 698}, 379 (2002).
\bibitem{Dani}
    P. Danielewicz {\it et al.}, Phys. Rev. Lett. {\bf 81}, 2438 (1998).
\bibitem{Sahu}
    P. K. Sahu and W. Cassing, Nucl. Phys. A {\bf 712}, 357 (2002).
\bibitem{Toneev}
    E.G. Nikonov, A.A. Shanenko and V.D. Toneev,
    Heavy Ion Phys. {\bf 8}, 89 (1998).
\end{thebibliography}
\end{document}